\documentclass[acmsmall,screen,nonacm]{acmart}
\renewcommand\footnotetextcopyrightpermission[1]{}
\usepackage[english]{babel}

\usepackage{tikz}

\usepackage{xcolor}
\usepackage{enumitem}
\usepackage{pbox}
\usepackage{multirow}
\usepackage{tabularx}
\usepackage{makecell}
\usepackage{mathtools}
\usepackage{caption}
\usepackage{float}
\usepackage{booktabs}
\usepackage{xurl}
\usepackage{gensymb}
\usepackage{subfigure}

\usepackage{hyperref}
\usepackage{graphicx}

\newcommand{\parab}[1]{\vspace{0.05in}\noindent\textbf{#1}}

\newcolumntype{L}[1]{>{\raggedright\let\newline\\\arraybackslash\hspace{0pt}}m{#1}}
\newcolumntype{C}[1]{>{\centering\let\newline\\\arraybackslash\hspace{0pt}}m{#1}}
\newcolumntype{R}[1]{>{\raggedleft\let\newline\\\arraybackslash\hspace{0pt}}m{#1}}

\usepackage{pifont}%

\renewcommand\footnotetextcopyrightpermission[1]{} %
\setcopyright{none}

\acmDOI{}

\acmISBN{}

\acmConference[]{}
\acmYear{}
\copyrightyear{}

\acmPrice{}

\begin{document}
\title[]{Characterizing the Role of Power Grids in Internet Resilience}

\author[]{Sangeetha Abdu Jyothi}
\affiliation{%
  \institution{UC Irvine, VMware Research}
}
\email{sangeetha.aj@uci.edu}

\begin{abstract}
Among critical infrastructures, power grids and communication infrastructure are identified as uniquely critical since they enable the operation of all other sectors. Due to their vital role, the research community has undertaken extensive efforts to understand the complex dynamics and resilience characteristics of these infrastructures, albeit independently. However, power and communication infrastructures are also interconnected, and the nature of the Internet's dependence on power grids is poorly understood. 

In this paper, we take the first step toward characterizing the role of power grids in Internet resilience by analyzing the overlap of global power and Internet infrastructures. We investigate the impact of power grid failures on Internet availability and find that nearly $65\%$ of the public Internet infrastructure components are concentrated in a few ($< 10$) power grid failure zones. More importantly, power grid dependencies severely limit the number of disjoint availability zones of cloud providers. When dependency on grids serving data center locations is taken into account, the number of isolated AWS Availability Zones reduces from 87 to 19. Building upon our findings, we develop NetWattZap, an Internet resilience analysis tool that generates power grid dependency-aware deployment suggestions for Internet infrastructure and application components, which can also take into account a wide variety of user requirements. 
\end{abstract}

\maketitle

\section{Introduction}
The geo-distributed Internet infrastructure comprises a multitude of components administered by a diverse set of entities. Autonomous Systems that form the backbone of the Internet are composed of high-end routers and long-haul links that span thousands of kilometers. Content providers host applications in hyperscale data centers with hundreds of thousands of servers. The Domain Name System with a hierarchy of servers and Content Delivery Networks with caches located close to users are some of the other key components of the vast Internet ecosystem. 

While the Internet itself is a sophisticated network with multiple layers and complex dynamics, it is not a standalone entity. The Internet depends on several other infrastructures. All hardware components that constitute the Internet rely on a power supply. Typically, the Internet components draw power from electric grids in the region in which they are installed. Recently, some infrastructures, such as hyperscale data centers, have partially or completely switched to renewable energy, which may be locally generated. However, routers and links on paths connecting to these data centers mostly continue to rely on regional power grids. Data centers also require cooling mechanisms, which often depend on local water supplies. Transportation infrastructure plays an important role during the deployment phase of Internet infrastructure components. Thus, the deployment and functioning of the Internet are intertwined with other critical infrastructures in our society. 

Among the sixteen designated critical infrastructures in the US, the Internet and power grids are identified as \textit{uniquely critical} due to the enabling function they provide for all other critical sectors~\cite{ppd}. Despite the pivotal role these infrastructures play in our society, some crucial aspects related to their resilience are overlooked today. The cascading impact of communication failure in interdependent power grids and networks used to send control signals to the grid has been studied before~\cite{power-comm1,power-comm2,power-comm3,power-comm4}. However, the dependence of the broader, global Internet on power infrastructures, which can lead to complex dynamics during large-scale failures, is poorly understood. 

In this paper, we take the first step towards characterizing the role of power grids in Internet resilience. Towards this goal, first, we examine the global power infrastructure and identify its features that are pertinent to Internet resilience. Second, using a broad range of power grid and Internet datasets, we analyze the dependence of the Internet on power infrastructures. Finally, we evaluate the impact of large-scale power grid failures on the Internet under high-impact failure scenarios---extreme weather events, solar superstorms, and manmade attacks such as Electromagnetic Pulse (EMP) attacks and cyber attacks.

Similar to the Internet, the global power infrastructure is also composed of multiple administrative domains. A Wide Area Synchronous Grid (WASG) is an electric grid that has a large geographic coverage and operates at a synchronized frequency. A WASG may be composed of multiple regional grids managed by different energy companies, which are electrically tied together with synchronized frequency. Grids around the globe are moving towards increased synchronization~\cite{geni,usgrid-info,eu-info,argentina-info,arab-info,india-info,sapp-info,russia-info} and, in turn, larger WASGs due to the benefits of interconnection such as lower cost and supply security. However, WASGs also increase grid vulnerability. 

WASGs are susceptible to cascading failures that can extend across the entire WASG. Several recent WASG outages~\cite{argUrug,euro06_1,india_1,usNE_1} that affected millions of people were caused by cascading failures triggered by small, local events. As nations move towards increased synchronization, the risk of such cascading failures also increases. More importantly, threats to power grids from a range of natural and manmade sources are also intensifying. Extreme weather events that are increasing in frequency and strength due to climate change were shown to increase the probability of cascading WASG failures by $30\times$~\cite{arizona-1}. Solar superstorms constitute another natural hazard that is expected to cause large-scale power outages lasting months to years~\cite{lloyds}. EMP attacks~\cite{emp-1,emp-2} and cyberattacks~\cite{cyberattack-1,cyberattack-2} are human-initiated threats capable of inducing large-scale power grid failures.

In light of vulnerabilities of the power infrastructure, it is imperative to understand its indirect impact on the Internet to improve the overall resilience of Internet-based applications. To enable this study, we assemble a broad set of power grid and Internet infrastructure datasets. We generate a WASG dataset by combining recent information from regional energy grid websites and the Global Energy Network Institute (GENI)~\cite{geni}. To study the impact on the public Internet, we use the Internet router topology from CAIDA~\cite{caida-itdk}, DNS root server locations~\cite{dnsRS}, global Internet Exchange Point (IXP) dataset~\cite{ixps}, and Internet user statistics~\cite{internetStats}. We use Amazon data centers~\cite{amazon-location} as a representative example for cloud providers. Using these datasets, we study the nature of overlap between global Internet and power infrastructures. We also evaluate the impact of power grid failures on the Internet by analyzing the availability of various infrastructure components and the loss of connectivity in the Internet topology during failures.

Our analysis shows that nearly $65\%$ of Internet infrastructure components (IXPs, routers, DNS root servers, etc.) are concentrated in a few WASGs ($<10$), thereby increasing the risk of large-scale Internet failures even when a small number of WASGs are affected. Analysis of regions of Amazon cloud shows that multiple regions are powered by the same WASG, particularly in North America. When dependency on grids serving data center locations is taken into account, the number of isolated Amazon Availability Zones reduces from 87 to 19. Thus, grid dependencies severely reduce the number of isolated cloud zones. 

Failure analysis under extreme weather events~\cite{climate-1,climate-2} and targeted attacks~\cite{emp-1,emp-2,cyberattack-1,cyberattack-2} that have regional impact show the highest fraction of Internet components being unreachable under European continental grid and North American Eastern interconnection failures. Failure analysis under solar storms with global impact shows that the fraction of Internet components affected will be higher than previous estimates~\cite{sigcomm21} due to the geographic spread of WASGs.

We develop NetWattZap, an Internet resilience analysis tool for generating resilience deployment suggestions for Internet infrastructure and application components that take into account power grid dependencies. NetWattZap is built atop maps of the global power grid and Internet infrastructure components. This tool will generate deployment suggestions that can satisfy a wide range of user requirements, including latency, cost, and various levels of resilience. Additionally, the tool is easily extensible to any user-defined linear constraints. We demonstrate the capabilities of NetWattZap using multiple use cases: finding grid failure resilient (i) data centers for a single user with latency constraints, (ii) data centers for a global application with geo-distributed users, and (iii) IXPs for large Autonomous Systems that want to pair at multiple locations.

Note that while the Internet relies on the power grid for energy, the power grid also relies on communication between various locations for its proper functioning. Although the dependencies are bidirectional, the focus of this paper is on understanding the dependence of the Internet on power grids. The challenges in the reverse direction have received more attention, particularly from the power infrastructure research community~\cite{power-comm1, power-comm2, power-comm3, power-comm4}.

\vspace{1mm}

\noindent
In summary, we make the following contributions:
\vspace{-2mm}

\begin{itemize}[leftmargin=*]
    \item We take the first step towards quantifying the impact of power grid failures on the Internet infrastructure by analyzing dependencies between the two infrastructures.
    \item Our analysis of overlap between power and communication infrastructure shows that $65\%$ of Internet infrastructure components are concentrated in fewer than 10 WASGs.
    \item We show that the number of Availability Zones of cloud providers (isolated locations used for fault tolerance) reduces drastically when power grid dependence is taken into account.
    \item Our analysis reveals that targeted attacks on the European continental grid and the North American Eastern interconnection can lead to the largest fraction of the Internet being unreachable.
    \item We show that the impact of solar storms on the Internet will be greater than previous estimates when the geographic spread of WASGs is taken into account.
    \item We develop and demonstrate multiple use cases of NetWattZap, an Internet resilience analysis tool that generates power grid dependency-aware deployment suggestions.
\end{itemize}

\section{Understanding the Infrastructure}

\subsection{Power Infrastructure} 
An electric grid interconnects the sources of power generation with the consumers and is composed of power stations, substations, and low- and high-voltage transmission lines. Since the generated electricity cannot be stored (in the vast majority of grids), the electricity generation typically has to match the demand. 

A Wide Area Synchronous Grid (WASG) is an interconnection of electric grids covering a large geographic area operating at a synchronized frequency. Interconnection of electric grids to form WASGs is increasingly common today due to a variety of reasons: pooling of reserve capacity across multiple grids to improve availability during peak loads and, in turn, reduces the peak generation capacity for individual grids, economic benefits, supply security arising from diversity of energy generation sources, etc. Although power grids have been in existence since the 1800s, increased synchronization is a recent phenomenon; all large grid synchronizations happened in the past two decades~\cite{usgrid-info, eu-info, india-info, china-info, sapp-info}. Today, large WASGs exist in all inhabited continents. For example, the continental US has three WASGs: the Eastern Interconnection, the Western Interconnection, and the Texas grid. The synchronous grid of continental Europe covers more than $20$ countries. 

In addition to the Alternating Current (AC) based WASGs, some regions also rely on high-voltage direct current (HVDC) transmission to transfer power between grids that are not synchronized. HVDC lines are typically employed between grids operating at different frequencies, between grids operating at the same frequency but with a phase shift, for transferring power from renewable energy sources such as off-shore windmills, etc. However, global deployment of HVDC lines is limited due to a high failure rate for equipment compared to AC systems and the need for additional expensive conversion equipment. While Ultra High-Voltage Direct-Current (UHVDC) is more promising, the current deployment of UHVDC lines is very limited due to exorbitant costs. Thus, AC-based WASGs are more common across the globe~\cite{unGrid}.

\subsection{Power Grid Resilience}
\label{subsec:power-res}
\parab{Power Grid Resilience Criteria.} Power grids are typically designed for ``N-1 security'', i.e., if a \textit{single} component in the grid fails, the grid is guaranteed to continue its operation without affecting electricity supply or causing cascading failures. The European continental grid~\cite{ucteTxn} and WASGs in the US~\cite{usTxn} cite N-1 security as an operation criterion. While N-k security requirements are well-understood, solutions that improve resiliency under more than one simultaneous failure are rarely adopted in practice due to the high costs involved~\cite{n-k-1,n-k-2}.

Power grid resilience is also affected by the complexity of WASGs. While WASGs offer great benefits of interconnection, they also introduce several challenges in the large-scale system. In addition to higher costs and technical complexities, WASGs increase the vulnerability of the grid. More specifically, the interconnection of multiple grids at synchronized frequency is susceptible to \textit{cascading failures that can bring down an entire WASG}~\cite{unGrid}. In other words, the entire region covered by a WASG could suffer from a power outage at the same time~\footnote{North American Electric Reliability Corporation (NERC) define cascading blackouts as ``the uncontrolled successive loss of system elements triggered by an incident at any location''~\cite{nercTerms}. Interruptions contained within a localized area are referred to as unplanned interruptions or disturbances. In this paper, our focus is on cascading blackouts.}. Analytical studies have established the increasing probability of cascading failures as grid sizes increase with extensive interconnections~\cite{cascade-1}. Studies have also shown that the failure of a very small subset of elements within a brief time window has the potential to cause large-scale cascading outages in today's grids~\cite{small-1,small-2}. In fact, modeling of cascading failures in large power grids is an active research area in the power grid community~\cite{topo-impact,topo-impact2,topo-impact3,cascade-model1,cascade-model2,cascade-model3,cascade-model4,cascade-model5,cascade-model6,cascade-model7,cascade-model8,cascade-model9}.

Currently, grids around the globe are moving towards massive WASGs with increasing synchronization efforts (continental-scale in many cases). Several efforts that will further increase the span of WASGs are in progress in Europe, Africa, South America, etc.~\cite{baltic-info,andean}. Although changes in political relations have led a few countries to leave WASGs in some cases, only a single instance of breaking up a synchronized grid for the purpose of improving resilience is known. China Southern Power Grid disconnected Yunnan province from the southern grid in China to improve resilience~\cite{china-1}. This was made possible by the significant deployment of expensive HVDC lines in the region. All countries except China are currently heading towards increased synchronization, and there are no publicly known efforts to break up WASGs (primarily due to the exorbitant costs). 

\parab{Real-World Large-Scale Grid Failures.} Real-world examples of WASG cascading failures are aplenty. Recently, on $15$ Aug 2023, a power outage affected all states in Brazil except one which was disconnected from the country grid~\cite{br-real-2023}, in turn causing widespread Internet outages (Appendix~\ref{sec:br-real}). On $16$ June $2019$, misconfigurations during the repair of a tower in Argentina led to a $14$-hour power outage across Argentina, Paraguay, and Uruguay for $48$ million people~\cite{argUrug}. The largest WASG power outages on $30$ and $31$ July $2012$ in India that affected over $400$ million people were triggered by the overloading of a single link and misoperation of its protection system~\cite{india_1, india_2}. The major $2006$ European blackout that affected millions of people was due to miscommunication between countries during a planned maintenance operation~\cite{euro06_1,euro06_2}. The cause for the blackout of 2003 across Northeastern states of the US that affected $55$ million people was identified as a software bug in the alarm system~\cite{usNE_1}. The blackouts listed above and several others were caused by cascading failures in a WASG triggered by small, local events.

\parab{Power Grid Threats.} With increasing grid synchronization and changing environmental and political conditions, several natural and manmade threats pose a risk for cascading power grid failures, and in turn, Internet failures. We highlight (a non-exhaustive set of) three significant threats---(i) increasing frequency of extreme weather events due to climate change, (ii) solar superstorms, and (iii) Electromagnetic Pulse (EMP) attacks. 

A broad range of scientific studies has concluded that climate change is causing an increase in the frequency and intensity of  hurricanes, floods, landslides, blizzards, and other extreme weather events around the globe~\cite{climate1, climate2, climate3, climate4}. The trend is expected to worsen in the near-future~\cite{climate5}. Climate change-related factors increase the probability of large-scale power outages through increased risk of mechanical failures~\cite{arizona-1}, inability to operate certain equipment under extreme weather conditions~\cite{texas-1}, etc. The likelihood of cascading failures is also predicted to rise due to increasing load and additional fluctuations in the grid caused by climate change~\cite{cascading-climate}. For instance, a $1\degree$C increase in ambient temperature was shown to increase the probability of cascading power outages by $30\times$ in Arizona~\cite{arizona-1}. 

Solar superstorms that cause geomagnetic disturbances are one of the largest natural threats to our power infrastructure. The collapse of the Quebec power grid during a moderate-scale solar storm in March $1989$ was triggered by seven different failures that happened on the grid within a time interval of $57$ seconds~\cite{tooImpToFail}. In its solar storm advisory~\cite{cisa-1}, the US Cybersecurity and Infrastructure Agency (CISA) pointed out that power grids in northern and coastal regions of the US would be severely affected by multi-point failures during powerful solar storms and that the currently known mitigation strategies are unlikely to be adopted in practice due to their exorbitant costs. Lloyd's report~\cite{lloyds} noted that large parts of the US could be without power for $1-2$ years in the event of a solar superstorm. The National Academy of Sciences workshop on severe space weather effects~\cite{NAP12507} also identified that the increased power transfer across grids today will worsen the impact of a solar storm, and power grids will experience damages of unprecedented proportions and, in turn, long-term blackouts. 

Electromagnetic Pulse (EMP) attacks (or High-altitude Electromagnetic Pulse (HEMP) attacks) constitute another class of high-impact, low-probability events that pose a threat to power grids. EMPs generate strong electromagnetic fields that produce a high voltage of thousands of volts on conductors, causing severe damage to power grid devices and control systems~\cite{emp-1}. While many components in power grids are placed in metallic boxes to mitigate the impact of external electromagnetic fields, these protection mechanisms are typically insufficient to withstand EMP attacks~\cite{emp-1}. The US commission that assessed the threat of EMP attacks reported that hostile state actors and terrorist organizations could obtain the greatest utility from nuclear weapons by using them for EMP attacks~\cite{emp-2}. The impact of EMP attacks was observed in the real world during controlled atmospheric nuclear testing conducted by the US and Soviet Union in the 1960s. During these tests, electrical equipment and even underground buried cables hundreds to thousands of kilometers away were damaged by EMP~\cite{emp-2}. 

It is worth noting that the three classes of threats listed above also pose a direct threat to the communication infrastructure, albeit to a lesser extent. Extreme weather events can inflict damage on Internet infrastructure~\cite{lightsOut}. Solar storms pose a threat to long-haul cables and satellite communications~\cite{sigcomm21}. Wireless and radio infrastructure is expected to suffer significant damage under EMP attacks, alongside less severe damage in the wired infrastructure~\cite{emp-2}. However, the focus of this paper is the indirect impact of power grid failures on the Internet.

\subsection{Infrastructure Interdependencies}
Power grids, water distribution, communication, healthcare, transportation, and other critical infrastructures  have complex interdependencies~\cite{interdep-1}. The strong interconnection between critical infrastructures in today's society poses a significant threat during large-scale failure scenarios. Hence, understanding these interdependencies is critical for improving infrastructure resilience and, in turn, ensuring societal stability during disaster scenarios. 

In addition to the risk of cascading failures within WASGs that constitute the power infrastructure, the nature of interconnection between power and other infrastructures also introduces the risk of cascading failures across multiple critical infrastructures, such as communication and water, during power grid failures. The bidirectional interdependence of communication and power infrastructure was noted in prior work~\cite{interdep-1}. Earlier efforts also looked at the two infrastructures' interdependence~\cite{heter-1,heter-2}, albeit on a much smaller scale. However, a detailed analysis of the impact of power grid failures on Internet infrastructure at today's scale is conspicuously absent. This paper aims to fill this gap.

\section{Datasets}
To understand the impact of large-scale failures of WASGs on the Internet infrastructure, we first compile a broad range of datasets that characterize the power grid and Internet infrastructures.

\begin{figure}[t]
	\centering
	\includegraphics[width=0.65\linewidth]{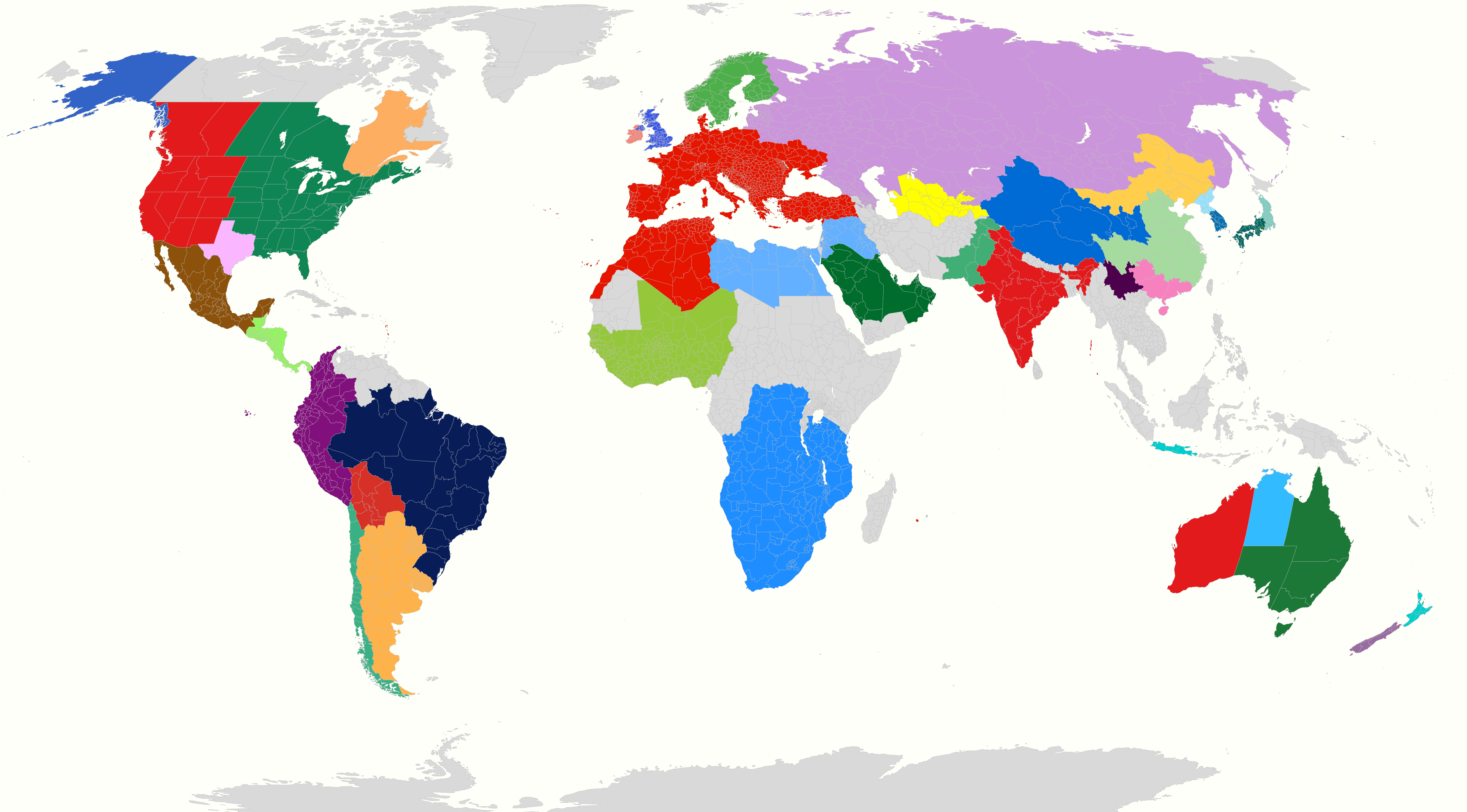}
	\caption{Map of $43$ Wide Area Synchronous Grids around the globe.}
	\label{fig:wasg-global}
\end{figure}

\begin{table}
\centering
\begin{tabular}{  l   l } 
    \textbf{Abbreviation} & \textbf{Info} \\
    \hline
    NA-E & North American Eastern Interconnection \\
    NA-W & North American Western Interconnection \\
    TX & US Texas grid \\
    MX & Mexican grid \\
    EU & Europe Continental Interconnection \\
    GB & Great Britain grid \\ 
    IN & Indian grid \\
    CN-1 & Chinese grid (North, Central, and East) \\
    CN-2 & Chinese grid (Southern) \\
    PK & Pakistan grid \\
    ID & Indonesia JAMALI grid \\
    KR & South Korea grid \\
    JP-1 & Japan (Eastern)\\
    Arab-1 & Arab grid (8 countries) \\ 
    UPS & Unified Power System (Soviet countries) \\
    BR & Brazil Grid \\
    SAPP & Southern African Power Pool \\    
    WAPP & West African Power Pool \\
    \hline
\end{tabular}
\vspace{1mm}
\caption{Abbreviations of top WASGs used in our study}
\label{table:wasg-info}
\end{table}

\subsection{Power Grid Infrastructure}
\label{subsec:powergrid-info}
Wide Area Synchronous Grids across the globe form the basis  of the power grid infrastructure. We generate the global WASG map (Figure~\ref{fig:wasg-global}) by combining recent information from regional energy grid websites and the Global Energy Network Institute (GENI)~\cite{geni}. In Table~\ref{table:wasg-info}, we give the abbreviations and a brief description of the top WASGs used in our analysis. Detailed information on WASGs is provided in Appendix~\ref{sec:wasg}. Note that the only other comprehensive global WASG map that was publicly available prior to this work was an image on Wikipedia generated in 2010~\cite{wasg-wiki}. This map did not reflect changes to WASGs across the globe in the past $12$ years and also had several errors. For example, missing WASGs that were formed in the past decade (e.g., WAPP), missing new countries added to existing WASGs (e.g., in Arab-1), etc.

\subsection{Internet Infrastructure}

\parab{Country-level Statistics:} 
We use statistics from the Internet World Stats website~\cite{internetStats} for estimating the users in various WASGs. This website provides information on the area, population, Internet users, and the percentage of Internet penetration for all countries across the globe. For WASGs with boundaries coinciding with country boundaries, accurate information for population and Internet users are directly available. For WASGs that do not fall in this category, we follow the methodology detailed in \S~\ref{subsubsec:methods} to obtain the statistics.

\parab{Internet Routers and Links:}
We obtain the Internet router and link topology information from CAIDA's Macroscopic Internet Topology Data Kit (ITDK)~\cite{caida-itdk}. We use the topology derived using MIDAR and iffinder \footnote{We find that this dataset contains a large number of non-routable IPs, with each such IP address assigned a unique node number. These non-routable IPs belong to the multicast address range 224.0.0.0-239.255.255.255. CAIDA annotates these as ``anonymous interfaces'' in the interface file but does not remove them from the nodes and links files. Hence, first, we perform data cleaning by removing non-routable IPs to obtain a dataset that only contains publicly routable addresses. All statistics are reported on the cleaned dataset.}. The CAIDA dataset contains over $52$ million (52 M) routers and $31.71$M links. Accurate geolocation is available for $51.24$ M routers (98.34\% of all routers). $0.86$ M routers (1.7\%) did not have a geolocation \footnote{CAIDA's node geolocation methodology involves two steps. First, each interface on a router is mapped to geolocation. Then, geolocation is assigned to the router only if all its interfaces map to the same location. In the absence of a perfect consensus among interface geolocations, the router is not assigned any location.}.

\parab{Internet Exchange Points (IXPs)}
We rely on the list of IXPs from the PCH Internet Exchange directory~\cite{ixps} for our analysis. This dataset contains $1097$ IXP locations across the globe, including their latitude and longitude information. 

\parab{DNS root servers}
We obtain the locations of root servers of the Domain Name System, which includes over $1000$ instances across $13$ root servers, from the root server directory~\cite{dnsRS}, including the geolocation of each server.

\parab{Data Center Locations}
We use Amazon~\cite{amazon-location} as a representative example of geo-distributed cloud infrastructures. Amazon Web Services (AWS) has one of the largest global footprints among cloud providers and the largest market share among cloud providers~\cite{amazon-stats}. The key location-based categorization of Amazon cloud includes \textit{Regions} that specify large geographic locations and \textit{Availability Zones} that denote isolated locations within each region. If replicas are placed across more than one availability zones within a region, Amazon guarantees availability even if one of them fails. Amazon has $27$ regions and $87$ availability zones around the globe. Recently, Amazon also introduced Local Zones within some Regions that allow users to place low-latency applications at specific locations closer to the users. Since Local Zones fall under the same WASG as the larger Region, we conduct our analysis at the scale of regions.

\section{Analyzing the Overlap}
We first analyze the overlap between WASGs and the Internet infrastructure components. Note that the focus of this section is solely on understanding the dependence of Internet on power grids. Failure analysis is deferred to the next section (\S~\ref{sec:failures}). 

\subsection{Methodology}
\label{subsubsec:methods}

\parab{WASG statistics:} Statistics for WASGs with boundaries coinciding with country boundaries are directly available from the Internet World Stats dataset. Most WASG boundaries overlap with country boundaries and their statistics can be directly estimated by aggregating country-level statistics. Only seven countries (the US, Canada, Japan, New Zealand, Australia, China, and Indonesia) have WASG boundaries that do not align with country boundaries; they coincide with state boundaries. For these WASGs, statistics at the state level are obtained from various sources~\cite{us-states,jp-states,nz-states,aus-states,china-states,id-states}. Accurate information is available for population and area at the state level for all these countries. For estimating Internet users at state granularity, the national average of Internet penetration obtained from Internet World Stats is used as the state-level penetration \footnote{Note that state-level Internet penetration could vary across states of a country based on geographical and demographic characteristics. However, due to a lack of better data sources, we rely on this method.}. 

\parab{Mapping Internet Components to WASGs:} The latitude and longitude information of various infrastructure components are available in the Internet infrastructure datasets. For Amazon data centers, we generate this information based on Region locations listed on the Amazon website~\cite{amazon-location}. 

For mapping Internet infrastructure geolocation to the corresponding WASG, we use the GeoPy library. GeoPy supports reverse geolocation of coordinates to an address that includes ISO 3166 codes for countries and states/provinces. ISO 3166 is the International Standard for country codes and codes for their subdivisions. Since latitude and longitude information of Internet infrastructure components are available, they are converted to ISO 3166 using this library. The decoded ISO country and state codes are then matched with the WASG database in which the member countries and states/provinces are stored in the same format.

\begin{figure*}[ht]
    \subfigure[Population Covered]{
    \centering
    \includegraphics[width=0.45\textwidth]{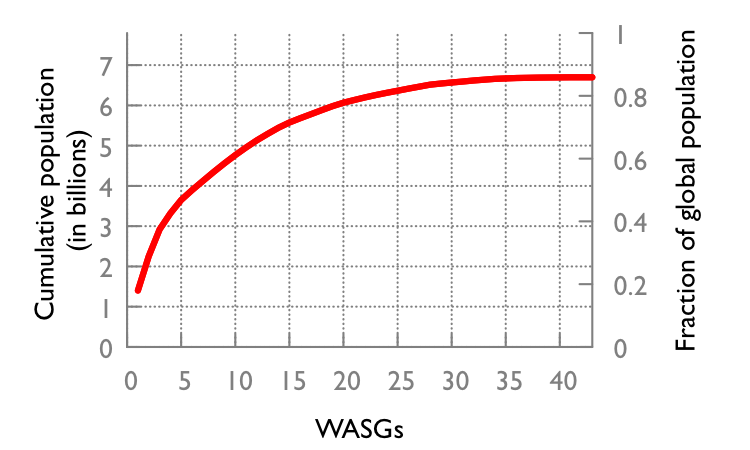}
    \label{fig:wasg-pop}
    }
    \subfigure[Internet users Covered]{
    \centering
    \includegraphics[width=0.45\textwidth]{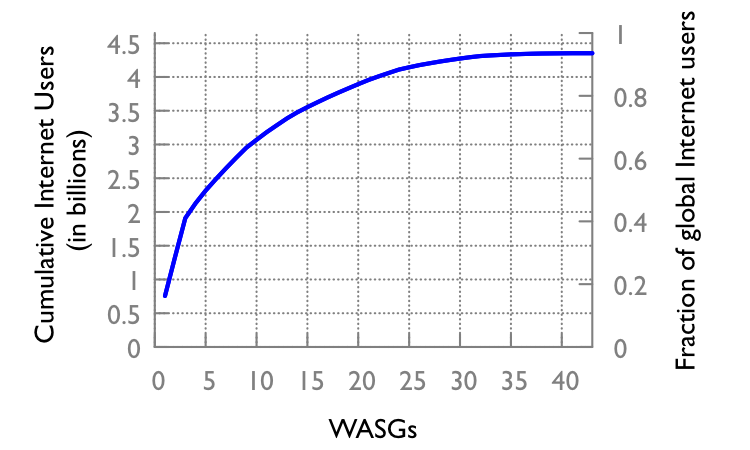}
    \label{fig:wasg-users}
    }
    \subfigure[Area Covered]{
    \centering
    \includegraphics[width=0.45\textwidth]{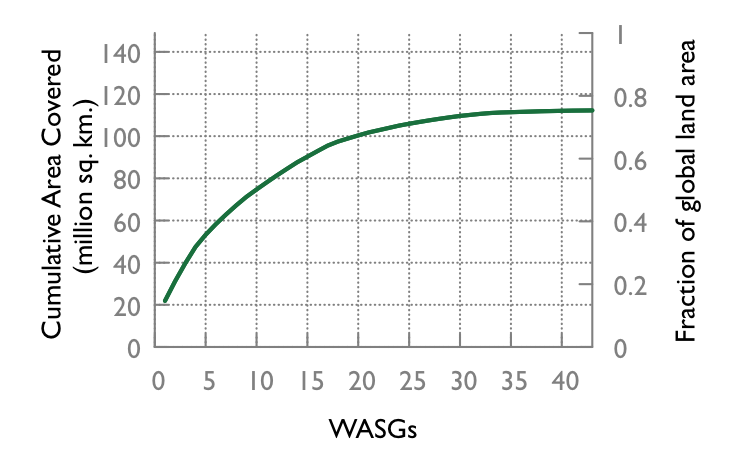}
    \label{fig:wasg-area}
    }
    \subfigure[IXPs Covered]{
    \centering
    \includegraphics[width=0.45\textwidth]{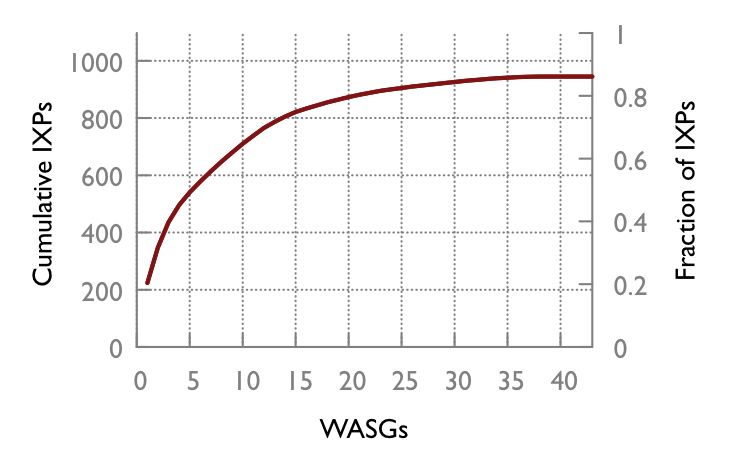}
    \label{fig:wasg-ixp}
    }
    \subfigure[Routers Covered]{
    \centering
    \includegraphics[width=0.45\textwidth]{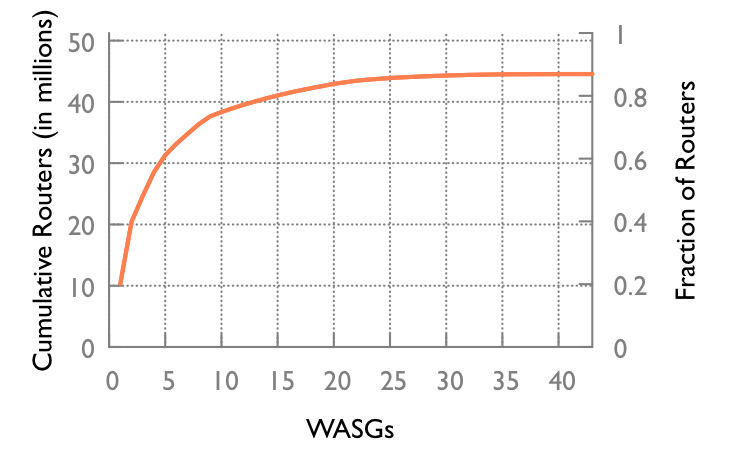}
    \label{fig:wasg-routers}
    }
    \subfigure[DNS Root Servers Covered]{
    \centering
    \includegraphics[width=0.45\textwidth]{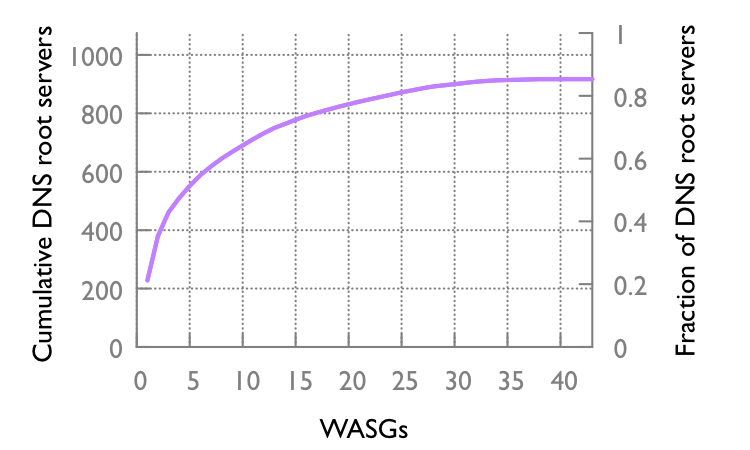}
    \label{fig:wasg-dns}
    }
\caption{Aggregate statistics of users and Internet infrastructure components across all WASGs. For each measurement, WASGs are sorted in decreasing order of size under the corresponding metric to highlight the impact of top contributor WASGs under each dimension.}
 \label{fig:wasg-stats}
\end{figure*}

\begin{figure*}[h!t]
\subfigure[Users in the WASG region]{
    \centering
    \includegraphics[width=0.45\textwidth]{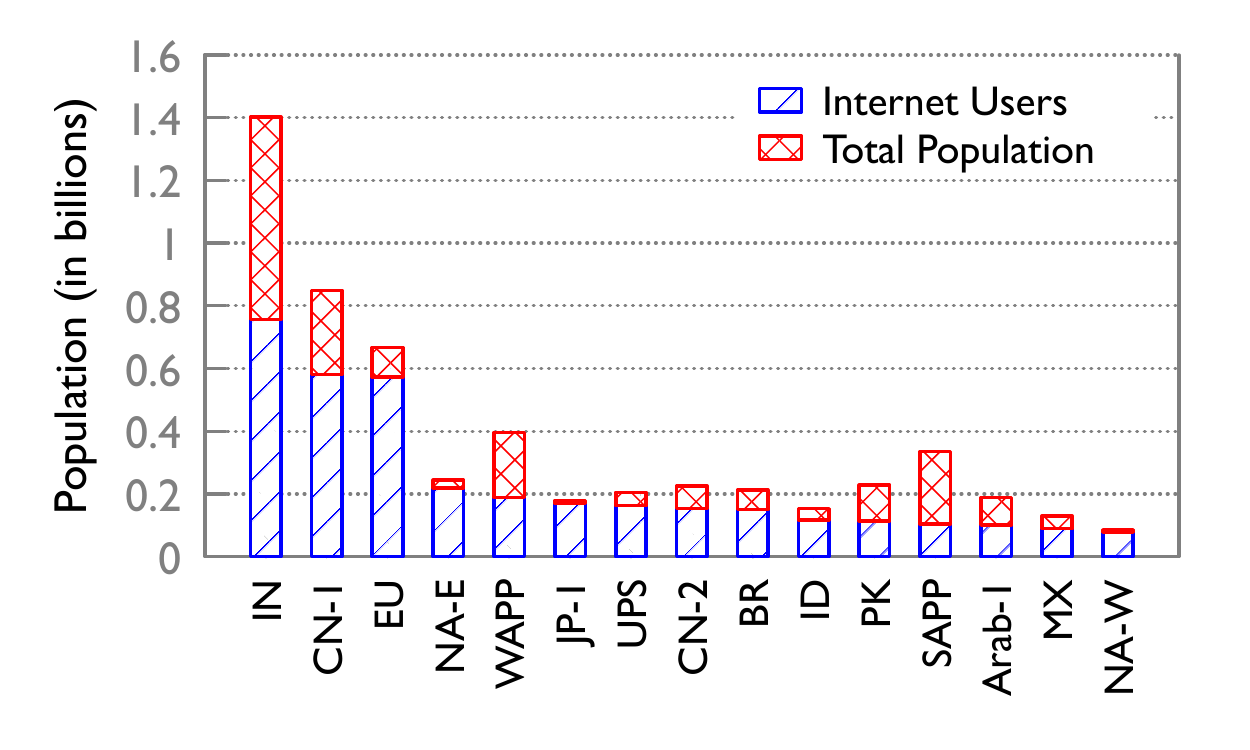}
    \label{fig:top15pop}
    }
    \subfigure[Area covered by WASG]{
    \centering
    \includegraphics[width=0.45\textwidth]{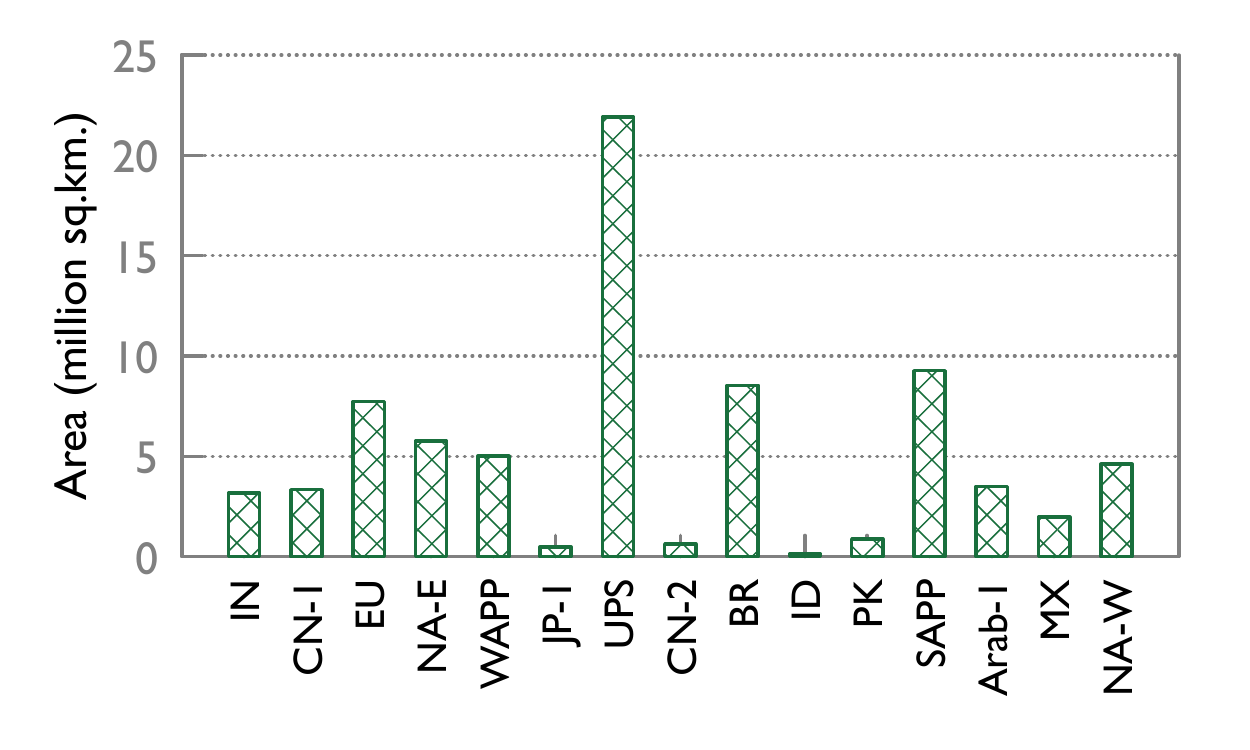}
    \label{fig:top15area}
    }
    \subfigure[IXPs in the WASG region]{
    \centering
    \includegraphics[width=0.45\textwidth]{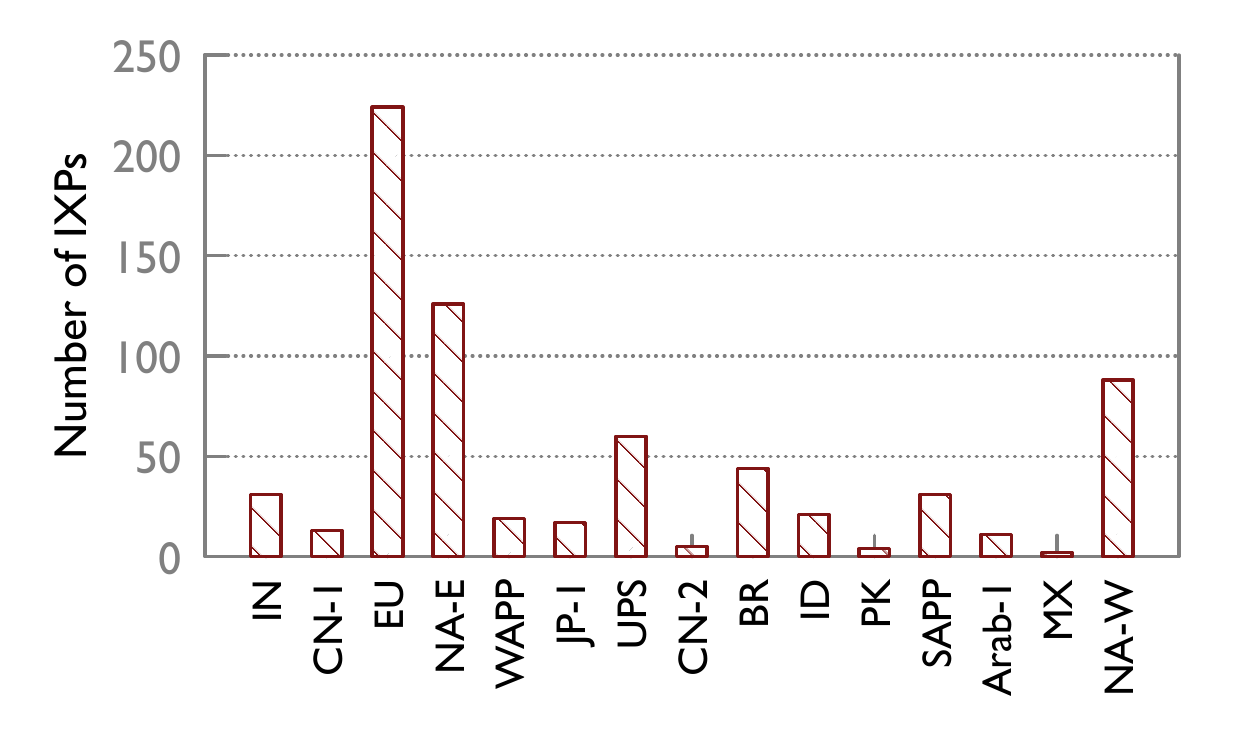}
    \label{fig:top15ixp}
    }
    \subfigure[Routers in the WASG region]{
    \centering
    \includegraphics[width=0.45\textwidth]{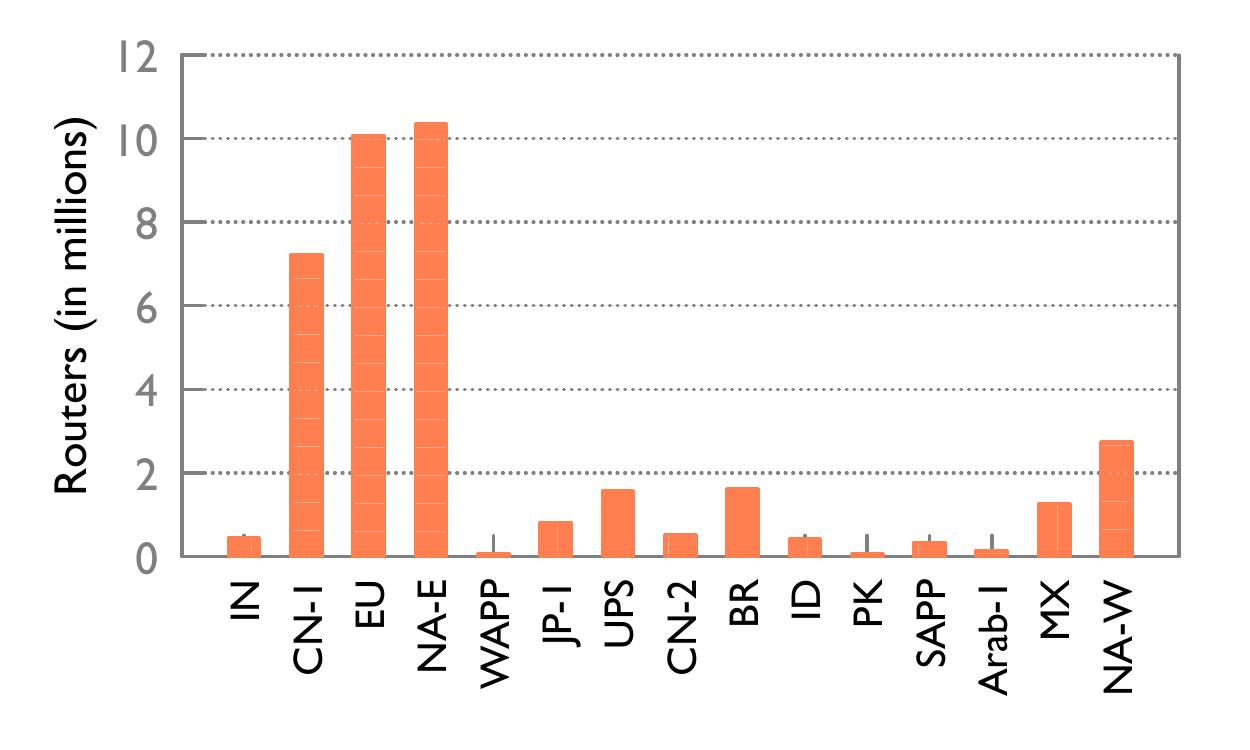}
    \label{fig:top15routers}
    }
    \subfigure[DNS root servers in the WASG region]{
    \centering
    \includegraphics[width=0.45\textwidth]{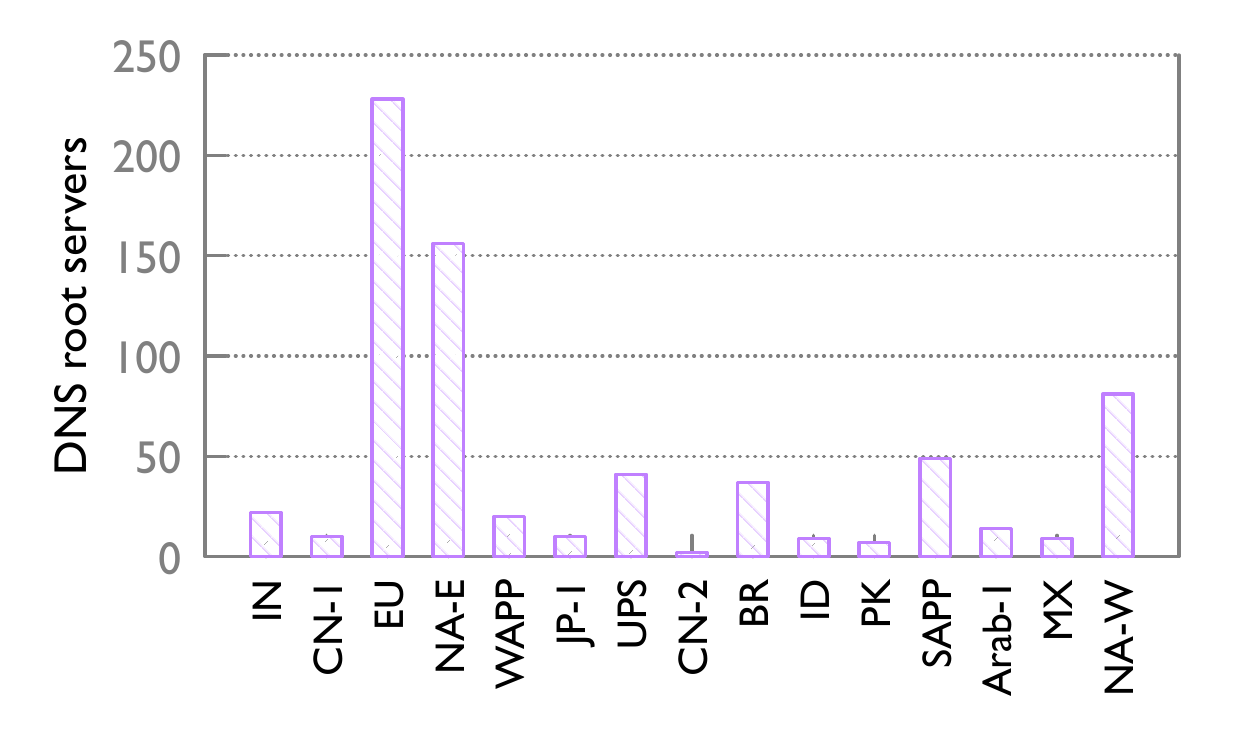}
    \label{fig:top15dns}
    }
    \subfigure[Amazon Regions and Availability Zones]{
    \centering
    \includegraphics[width=0.45\textwidth]{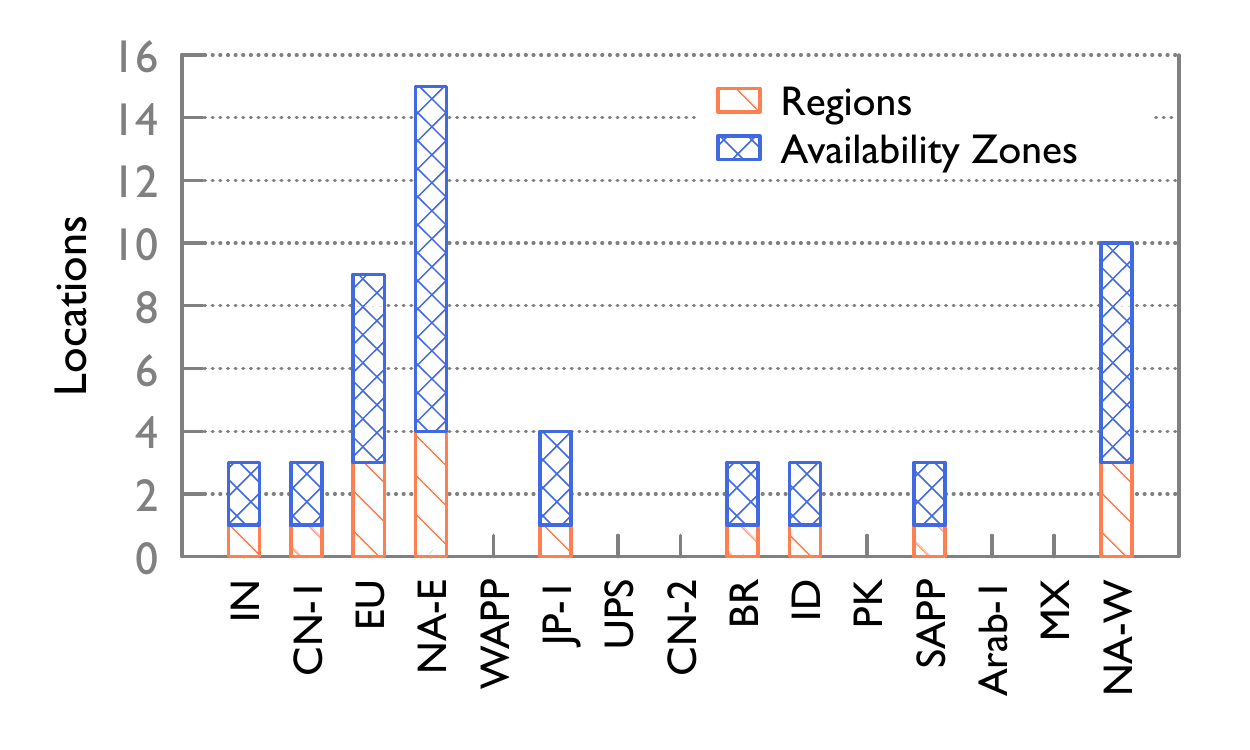}
    \label{fig:top15amazon}
    }
\caption{User and Internet infrastructure component distribution across top 15 WASGs. The top 15 WASGs are determined based on the number of Internet users in the WASG region and are maintained in the same order across all plots.}
 \label{fig:top15}
\end{figure*}

\subsection{Results}
\subsubsection{User and Infrastructure Statistics} \hfill\\
In Figure~\ref{fig:wasg-stats}, we present the aggregate statistics over $43$ WASGs. For analysis under each metric, WASGs are sorted in the decreasing order of value under that metric.

\parab{User distribution:} In Figures~\ref{fig:wasg-pop} and \ref{fig:wasg-users}, we observe that nearly $87\%$ of the world's population and $95\%$ of Internet users receive their energy supply from Wide Area Synchronous Grids. The six largest WASGs alone serve half of the world's population as well as Internet users. The largest grids in terms of users supported include the Indian grid, the interconnection of three regional Chinese grids, the grid of continental Europe, and the Eastern interconnection in North America. WASGs also extend over a significant fraction of today's land area, as shown in Figure~\ref{fig:wasg-area}.

\parab{Infrastructure component distribution:} In Figure~\ref{fig:wasg-ixp}, we evaluate the distribution of $1097$ Internet Exchange Points (IXPs) across the $43$ WASGs. $88\%$ of IXPs are powered by Wide Area Synchronous Grids. Three grids---the grid of continental Europe and the Eastern and Western interconnections in North America---together power $40\%$ of all IXPs. Analysis of Internet routers in Figure~\ref{fig:wasg-routers} also shows a heavily skewed distribution, with more than half of routers located in just three WASG regions. Nearly $40\%$ of Internet routers are located in the Eastern interconnection in North America and the continental European grid.

DNS root servers (Figure~\ref{fig:wasg-dns}), although skewed, have a more global spread. For example, the top five WASGs under this metric are the continental European grid, the Eastern and Western interconnections in North America, the Southern African Power Pool, and UPS which includes several former Soviet countries.

\subsubsection{Deep Dive into Top-15 WASG Characteristics}\hfill\\
The infrastructure component distribution analysis does not reveal fine-grained insights about the various metrics and their correlations. Hence, we conduct a deep dive into statistics of the top $15$ WASGs determined based on the Internet user population in the region. We rely on a population-based metric to determine the top WASGs since the complexity of the grid is correlated with the number of users supported. Analysis of only $15$ WASGs is plotted here due to space constraints, statistics for all $43$ WASGs are available in our dataset.

\parab{User statistics:} In Figure~\ref{fig:top15pop}, we plot the population and Internet user distribution of the top 15 WASGs. The codes for the top 15 WASGs and others used in the evaluation are available in \S~\ref{subsec:powergrid-info}. We observe in Figure~\ref{fig:top15pop} that Internet penetration differs widely across the top WASGs.

\parab{Area coverage:} In Figure~\ref{fig:top15area}, we observe that the area of a WASG has a low correlation with population/Internet users as well as Internet infrastructure component distribution. The largest WASG (UPS) in terms of area, for example, covers a sparsely populated region of Siberia with very limited users, whereas several densely populated small Asian countries feature among the top WASGs.

\parab{IXP distribution:} As shown in Figure~\ref{fig:top15ixp}, the continental grid of Europe has a high concentration of IXPs, nearly $2\times$ more than the next largest IXP hosting WASG, the North American Eastern Interconnection. In addition, while the continental grid hosts $224$ IXPs, the Nordic countries, Great Britain, and Ireland together have only $71$ IXPs ($38$, $28$, and $5$, respectively). North America, on the other hand, has a better distribution of IXPs across three WASGs, with $124$ in Eastern interconnection, $88$ in Western Interconnection, and 27 in the Texas grid region.

\parab{Router distribution:} Router distribution across WASGs (Figure~\ref{fig:top15routers}) differs significantly from IXPs. The North American Eastern Interconnection hosts a slightly higher share of routers compared to the European grid. While NA-E is home to over $10$M routers, NA-W only contains $2.75$M. India has substantially fewer routers in relation to its area and population density compared to other regions of similar characteristics in spite of a large Internet user population. This could be due to the poor coverage of routers in this region by the CAIDA dataset. 

\parab{DNS root server distribution:} DNS root server distribution (Figure~\ref{fig:top15dns}), although skewed, shows a broad prevalence of root servers. While European and North American grids host the largest fraction of root servers, the Southern African Power Pool also features in the top five root server regions.

\parab{Data Center Characteristics:} Amazon data center distribution is shown in Figure~\ref{fig:top15amazon}. Although AWS offers multiple Availability Zones within each Region, which are isolated, all Availability Zones in a Region share the same WASG. Hence, from the perspective of WASG failures, there is only a single ``Availability Zone'' in a WASG. Similarly, multiple Regions served by the same WASG also share a failure zone. 

In a detailed analysis of all AWS locations based on WASG boundaries, we observe that Amazon's $27$ Regions and $87$ Availability Zones reduced to $19$ Regions/Availability Zones which do not share the same WASG ($10$ in Asia,  $4$ in Europe, $2$ in North America, and $1$ each in South America, South Africa, and Australia). While North American Amazon Regions are limited to Eastern and Western Interconnections, Regions in Asia have a better spread across different countries. Even within China, the three locations (Beijing, Ningxia, and Hong Kong) are served by disjoint grids, thereby improving resilience under grid failures. Availability in North America can be improved by adding Regions in other disconnected grids, such as Texas, Quebec, or Mexico. 

\subsubsection{IP links and WASGs}\hfill\\
To understand the nature of overlap between WASGs and the Internet network topology, we map the endpoint routers of each IP link to WASGs and analyze the resulting distribution. Some routers (endpoints) may be located in regions without a WASG. For example, Somalia, Bhutan, etc., are not part of any WASGs. When both ends of an IP link are mapped to WASGs, these WASGs may be the same or different. Based on these observations, in Table~\ref{table:ip}, we show the distribution of links across various categories. We observe that only $4\%$ of the links ($1.3$M) have no WASG mapping at both endpoints. 96\% of links are powered by a WASG on at least one of its ends, with 33.7\% (10.7M) having a WASG mapping at both ends.

\begin{table}
\centering
\begin{tabular}{  l  r } 
\hline
 \textbf{Category} & \textbf{\# Links }  \\ 
 \hline
(i) Both endpoints mapped to WASGs  & 10.69 M \\ 
(ii) Only one endpoint mapped to a WASG & 19.71 M \\
(iii) None of the endpoints mapped to a WASG & 1.3 M  \\
 \hline
\textbf{\textit{Total}} & \textbf{31.71 M} \\
\hline
\end{tabular}
\vspace{1mm}
\caption{Categorization of IP links based on WASG mapping of endpoints}
\label{table:ip}
\end{table}

\begin{table}[]
    \parbox{.45\columnwidth}{
    \centering
    \begin{tabular}{ p{0.25\textwidth}   r }
    \hline
    \textbf{WASG pair} & \textbf{ \# IP links} \\
    \hline
        NA-E, \quad EU & 1661 K \\
        NA-E, \quad CN-1 & 856 K \\
        NA-E, \quad NA-E & 608 K \\
        CN-1, \quad EU  & 420 K \\
        EU, \qquad KR & 399 K \\
        NA-W, \quad EU & 379 K \\
        NA-E, \quad GB & 354 K \\
        NA-E, \quad NA-W & 351 K \\
        NA-E, \quad KR & 267 K \\
        NA-E, \quad UPS & 239 K \\
    \hline
    \end{tabular}
    \vspace{1mm}
    \caption{Top 10 WASG pairs based on IP links with geolocation at both endpoints (category (i) in Table~\ref{table:ip})}
    \label{table:wasg-pairs}
    }
    \hfill
     \parbox{.45\columnwidth}{
    \centering
    \begin{tabular}{ p{0.1\textwidth}   r }
    \hline
    \textbf{WASG} & \textbf{\# IP links} \\
    \hline
        NA-E & 5133 K \\
        EU & 3848 K \\
        CN-1 & 2881 K \\
        KR & 1558 K \\
        NA-W & 1265 K \\
        GB & 839 K \\
        UPS & 617 K \\
        BR & 486 K \\
        MX & 390 K \\
        TX & 386 K \\
    \hline
    \end{tabular}
        \vspace{1mm}
        \caption{Top 10 WASGs based on IP links with geolocation at one endpoint only (category (ii) in Table~\ref{table:ip})}
        \label{table:wasg-oneEnd}
    }
\end{table}

We analyze the links with geolocation information on at least one endpoint to understand the nature of overlap between power and Internet infrastructure topologies. In Table~\ref{table:wasg-pairs}, we list the top $10$ WASG pairs based on inter-WASG IP links. For this, we focus on IP links with geolocation on both endpoints of the link (in categories (i) of Table~\ref{table:ip}). The largest fraction of geolocated IP links is between the Eastern Interconnection of North America and the continental European grid. NA-E and EU also have the largest aggregate number of links to all other WASGs.

In Table~\ref{table:wasg-oneEnd}, we present the results of the analysis of links in category (ii), where only one of the link endpoints has a WASG mapping. Note that this category is much larger (in Table~\ref{table:ip}). NA-E and EU top the list in this set of links as well.

Note that in addition to missing geolocation information for known nodes in the CAIDA dataset, all nodes and links in the Internet are currently not visible in our datasets due to a limited number of vantage points. Moreover, there could be errors in geolocation and inference techniques used to generate the Internet topology dataset. In spite of these shortcomings, the CAIDA dataset provides the most comprehensive picture of the Internet infrastructure today. Hence, given the current lack of understanding of the nature of overlap between two complex infrastructures, we strive to analyze their interdependencies based on the best available datasets. When better Internet datasets become available, the accuracy of this analysis can be improved further in the future.

\section{Failure Analysis}
\label{sec:failures}

In this section, we present the evaluation of Internet infrastructure unavailability under large-scale WASG failures. We simulate the impact of two categories of threats: (i) regional impact events, such as climate change-induced extreme weather events and targeted human-initiated attacks, and (ii) global impact events, such as solar superstorms.

\subsection{Methodology}
We analyze the fraction of Internet infrastructure components that are unavailable under a variety of failure scenarios. The fraction of Internet users, IXPs, routers, DNS root servers, and Amazon regions that become unreachable are estimated using the geolocation of infrastructure components.

\subsubsection{Failure Zones. } Modeling of cascading failures is a topic that has garnered considerable attention in the power grid community, evidenced by a robust body of research~\cite{topo-impact,topo-impact2,topo-impact3,cascade-model1,cascade-model2,cascade-model3,cascade-model4,cascade-model5,cascade-model6,cascade-model7,cascade-model8,cascade-model9}. The extent of the impact caused by a cascading failure is influenced by a myriad of factors, such as the grid topology, the number of elements that fail simultaneously within a short interval, the load conditions during the failure event, and the transient dynamics of the entire system. A cascading failure might encompass an entire grid or only parts of it based on the WASG characteristics and the failure event. 

In our analysis, we treat an entire WASG as a unified failure zone. This approach aligns with the notion of failure zone in the power grid resilience research community~\cite{topo-impact,topo-impact2,topo-impact3,cascade-model1,cascade-model2,cascade-model3,cascade-model4,cascade-model5,cascade-model6,cascade-model7,cascade-model8,cascade-model9}, and is based on the understanding that the behavior of synchronized grid systems inherently interconnects their failure probabilities. It is guided by the fact that regions served by the same synchronized grid inherently possess a non-zero risk of experiencing joint failures.

Note that this definition also aligns with the definition of availability zones of cloud providers, which represent isolated failure domains. Some private Wide Area Networks also slice their network into smaller fault domains, each managed by a separate controller~\cite{blastshield}, to support network applications such as traffic engineering. In scenarios where these dependent networked systems need to factor in the resilience of the underlying power grid infrastructure, the notion of fault isolation needs to align between the two infrastructures. 

Hence, we characterize the impact of regional and global failure events by treating each WASG as an isolated failure zone.

\subsubsection{Connectivity Analysis.} We estimate the loss of connectivity in the Internet topology during grid failures by generating a graph with WASGs as nodes. We use the NetworkX package~\cite{networkx} for our graph evaluations. Note that the CAIDA topology dataset does not include the capacity of IP links. Hence, the total capacity of Internet links between various WASGs is not known. We use the number of IP links with endpoints at two different WASGs as a proxy for capacity between the two WASGs. Thus, we create an undirected graph with nodes as WASGs and capacity on links as the number of IP links between the endpoint WASGs. 

We first evaluated common graph metrics such as degree centrality and betweenness centrality on the WASG graph. However, we found these metrics to be less insightful due to the high interconnectivity in this graph. The graph of WASGs with $43$ nodes has an average degree per node of $36.8$. Since the vast majority of nodes have a link to other nodes, metrics such as betweenness centrality do not help in understanding connectivity patterns under failures. Hence, we use the maximum flow between all node pairs (WASG pairs) in the WASG graph to characterize the extent of connectivity. Since the link weights are the number of IP links between the connected WASGs, the magnitude of flow reveals insights about the connectivity. To understand the impact of node failures, we remove the failed nodes (WASGs) from the graph and recalculate the maximum flow for all pairs. We report the mean reduction in maximum flow for various failure scenarios.

The complexity of finding maximum flow between a pair of nodes in a graph with $V$ nodes and $E$ edges is $O(VE^2)$ using the popular Edmonds-Karp algorithm~\cite{ed-karp}. Repeating this computation for every node pair before and after failures is computationally expensive. Recall that most WASGs have IP links to all other WASGs.  Hence, we use the Gomory-Hu tree~\cite{gomory} for maximum flow computations across all pairs of nodes. The Gomory-Hu tree of an undirected graph with capacities on its links is a weighted tree that represents the minimum s-t cuts for all s-t pairs in the graph. Computing maximum flow between all pairs of nodes requires $V* (V-1)/2$ maximum flow computations, whereas constructing a Gomory-Hu tree requires only $V-1$ maximum flow computations. 

Note that the maximum flow evaluation on the WASG graph is only a proxy for the impact on end-to-end connectivity. Currently, the Internet measurement literature lacks more pragmatic metrics to appraise the end-to-end connectivity attributes of Internet topology beyond common graph metrics like centrality, which, unfortunately, offered limited insights in this specific context.

\begin{figure*}[ht]
    \subfigure[ EU]{
    \centering
    \includegraphics[width=0.45\textwidth]{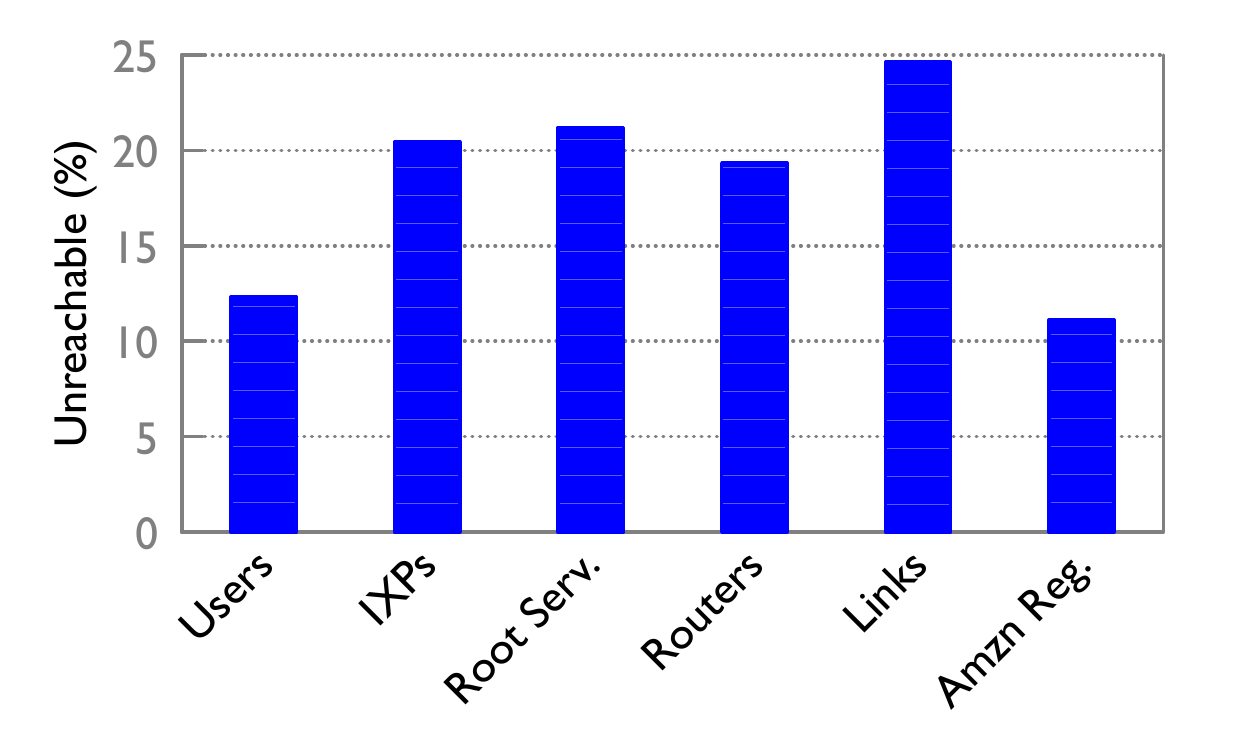}
    \label{fig:failure-eu}
    }
    \subfigure[NA-E]{
    \centering
    \includegraphics[width=0.45\textwidth]{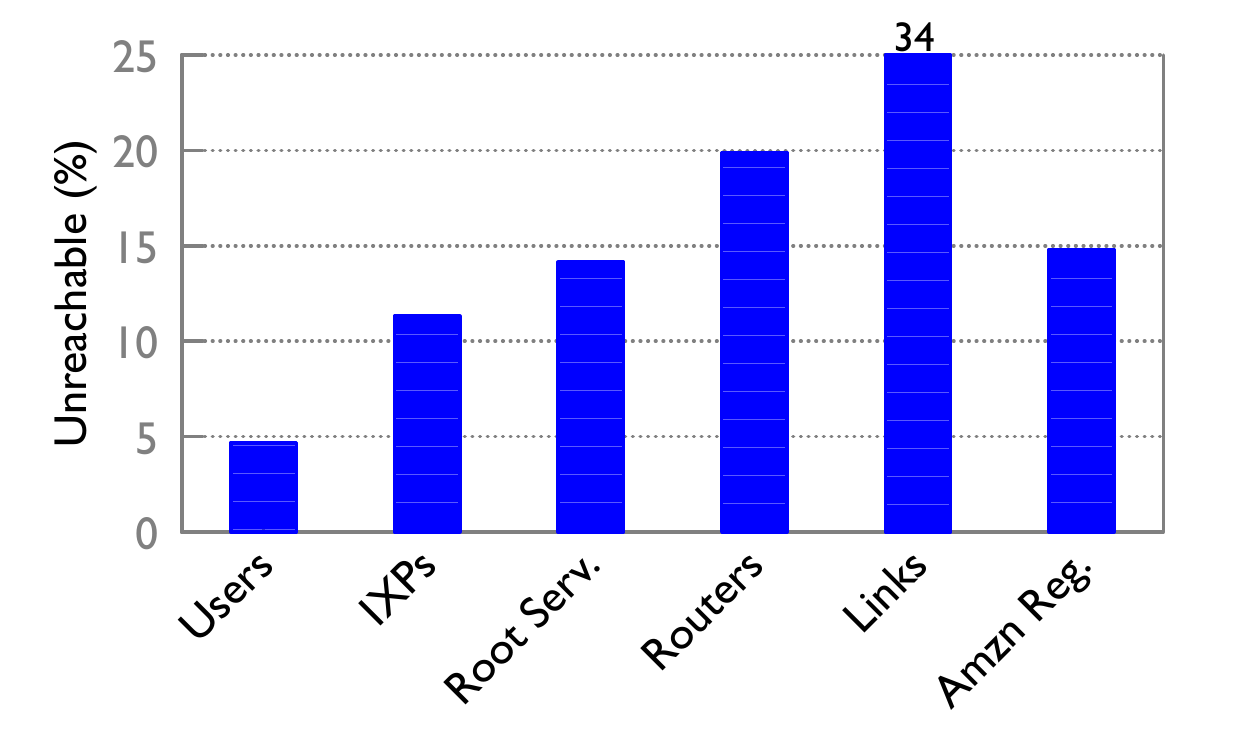}
    \label{fig:failure-na-e}
    }
    \subfigure[ NA-W]{
    \centering
    \includegraphics[width=0.45\textwidth]{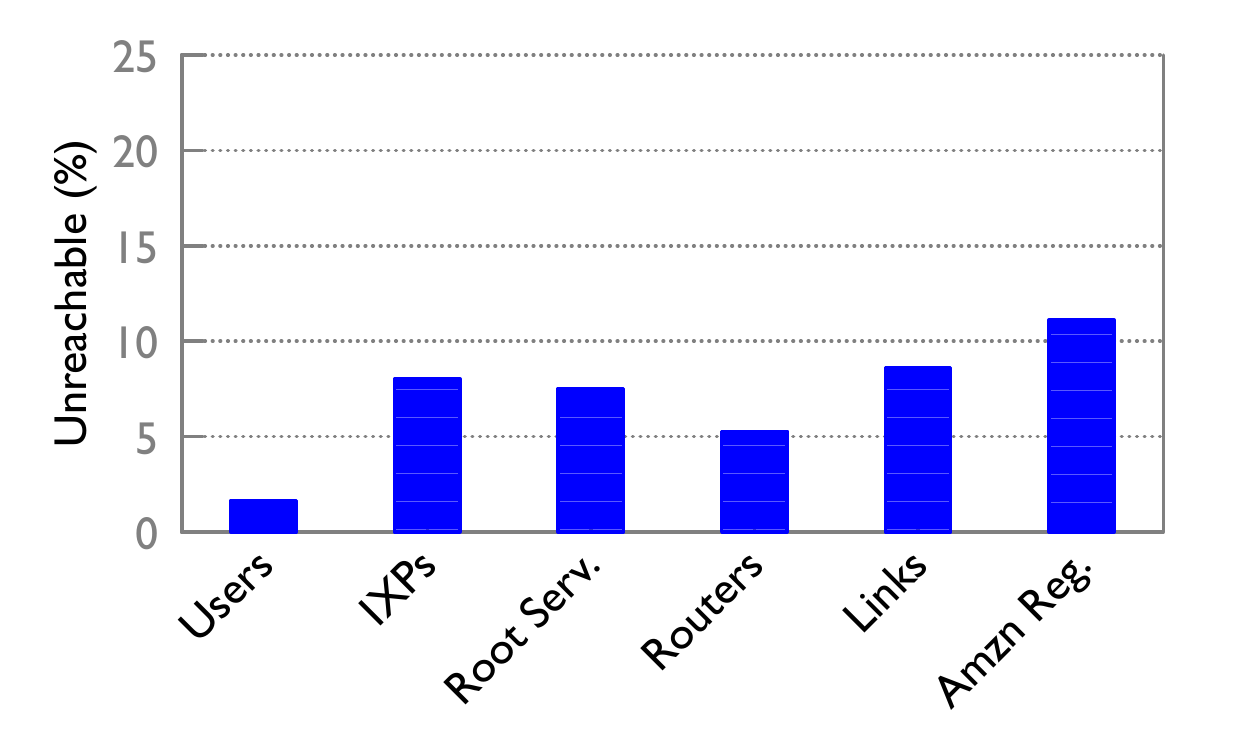}
    \label{fig:failure-na-w}
    }
    \subfigure[CN-1]{
    \centering
    \includegraphics[width=0.45\textwidth]{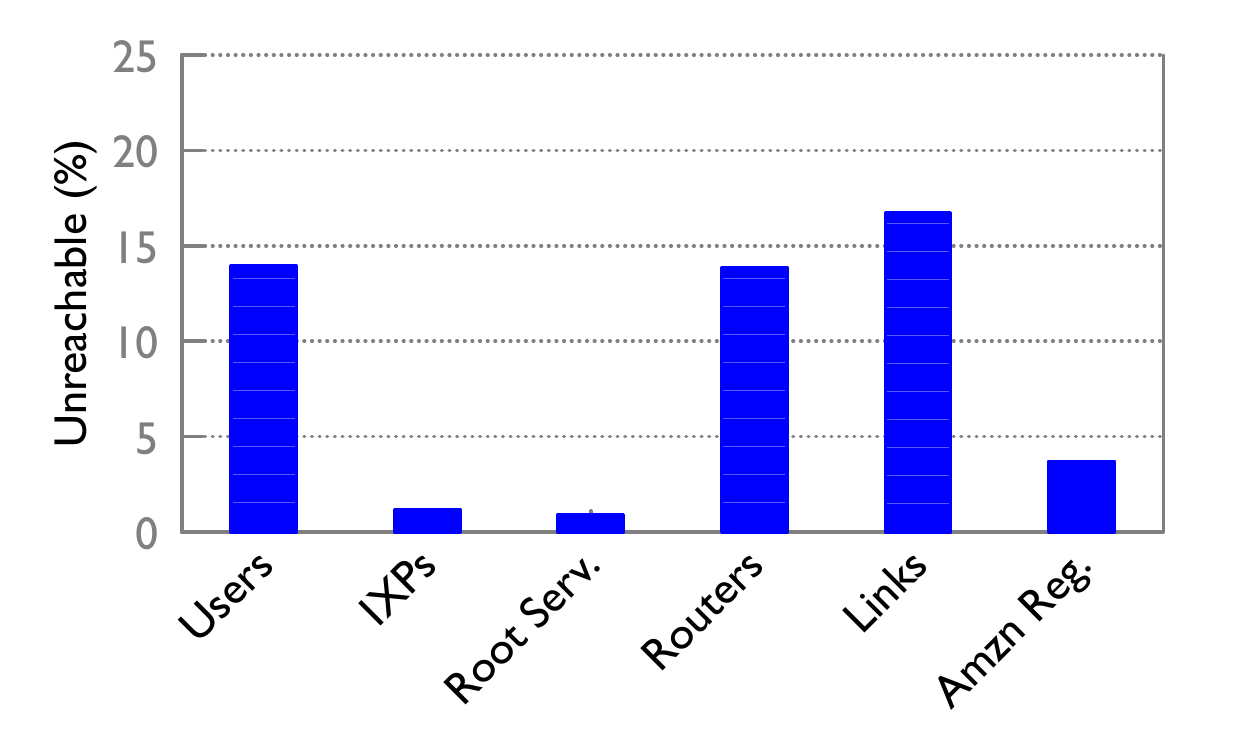}
    \label{fig:failure-cn-1}
    }
    \subfigure[UPS]{
    \centering
    \includegraphics[width=0.45\textwidth]{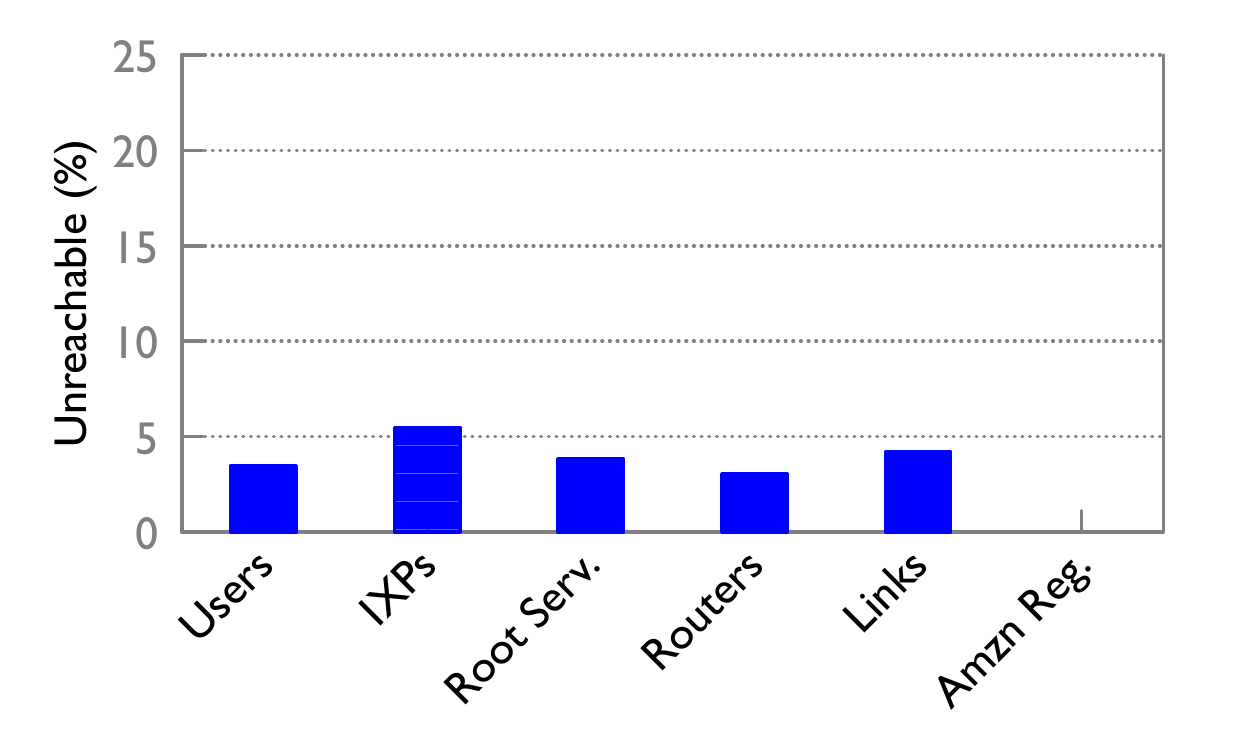}
    \label{fig:failure-ups}
    }
    \subfigure[BR]{
    \centering
    \includegraphics[width=0.45\textwidth]{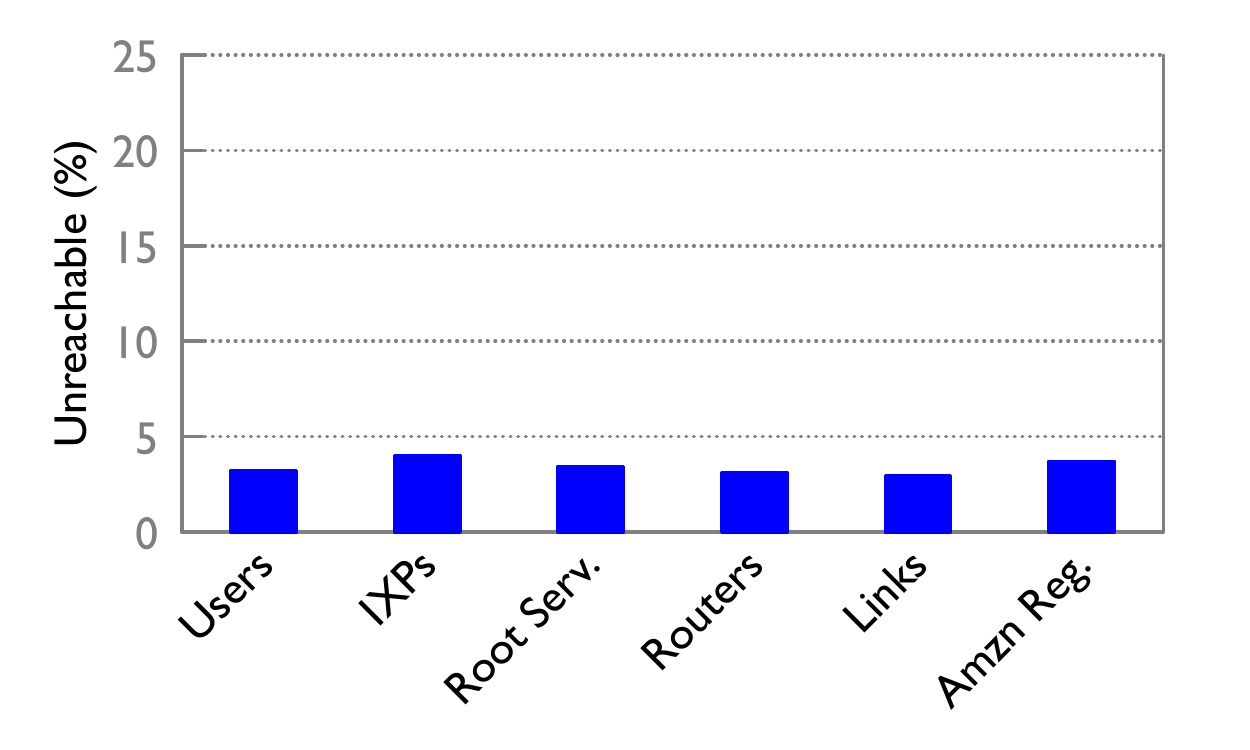}
    \label{fig:failure-br}
    }
\caption{Distribution of users and Internet infrastructure components that become unreachable under failures caused by extreme weather events or human-initiated attacks in various regions.}
 \label{fig:failure-regions}
\end{figure*}

\subsection{Regional Failures: Extreme Weather Events and Targeted Attacks}

Extreme weather events and targeted human-initiated attacks such as Electromagnetic Pulse (EMP) attacks and cyberattacks are capable of causing regional grid failures.

Data centers and other Internet infrastructure components such as IXPs generally have backup power for 24-48 hours using a wide variety of technologies such as UPS, diesel generators, batteries, etc.~\cite{dc-backup}. Hence, these infrastructures are immune to localized and short-lived power outages. We are interested in evaluating extreme weather events which are capable of causing power outages across large areas. Failure of WASG in the region can leave a data center unreachable even if it has backup power, and routers and links along the access path do not have a power supply.

Recent extreme weather events revealed unforeseen effects on the Internet infrastructure. In Europe, during the heat wave of the Summer of 2022, smaller cloud locations were forced to shut down equipment to prevent damage~\cite{turnoff-uk}. Google and Oracle experienced failure of cooling systems in their London data centers~\cite{goog-orac-heatwave}. Google also reported elevated error rates, latencies, and service unavailability in europe-west2 region~\cite{google-heatwave}. During the same heat wave, power production was also struggling due to the lower efficiency of power infrastructure components at high temperatures~\cite{heatwave-grid}. Similarly, EMP attacks~\cite{emp-1,emp-2} and cyberattacks~\cite{cyberattack-1,cyberattack-2} are also capable of bringing down regional grids. Hence, we evaluate the indirect impact on Internet infrastructure due to large-scale attacks on regional power grids.

In Figure~\ref{fig:failure-regions}, we analyze the impact of the failure of six large and significant grids spread across multiple continents (EU, NA-E, NA-W, CN-1, UPS, and BR). We report the percentage of users and Internet infrastructure components (IXPs, DNS root servers, routers, links, and Amazon AWS cloud regions) that become unavailable during regional WASG failures. We observe that failure of the EU grid will result in the largest fraction of IXPs and DNS root servers being unreachable, $20\%$ and $21\%$, respectively. In addition, each of EU and NA-E failure will leave nearly $20\%$ of Internet routers without power, but NA-E failure will affect more links ($34\%$ for NA-E vs. $24\%$ for EU). AWS cloud will lose $10\%$ of its regions if any of EU, NA-E, or NA-W fails. The impact of grid failures of UPS and BR is less severe.

\begin{figure}[h]
\centering
    \subfigure[Latitude Threshold of $40^{o}$]{
    \centering
    \includegraphics[width=0.45\textwidth]{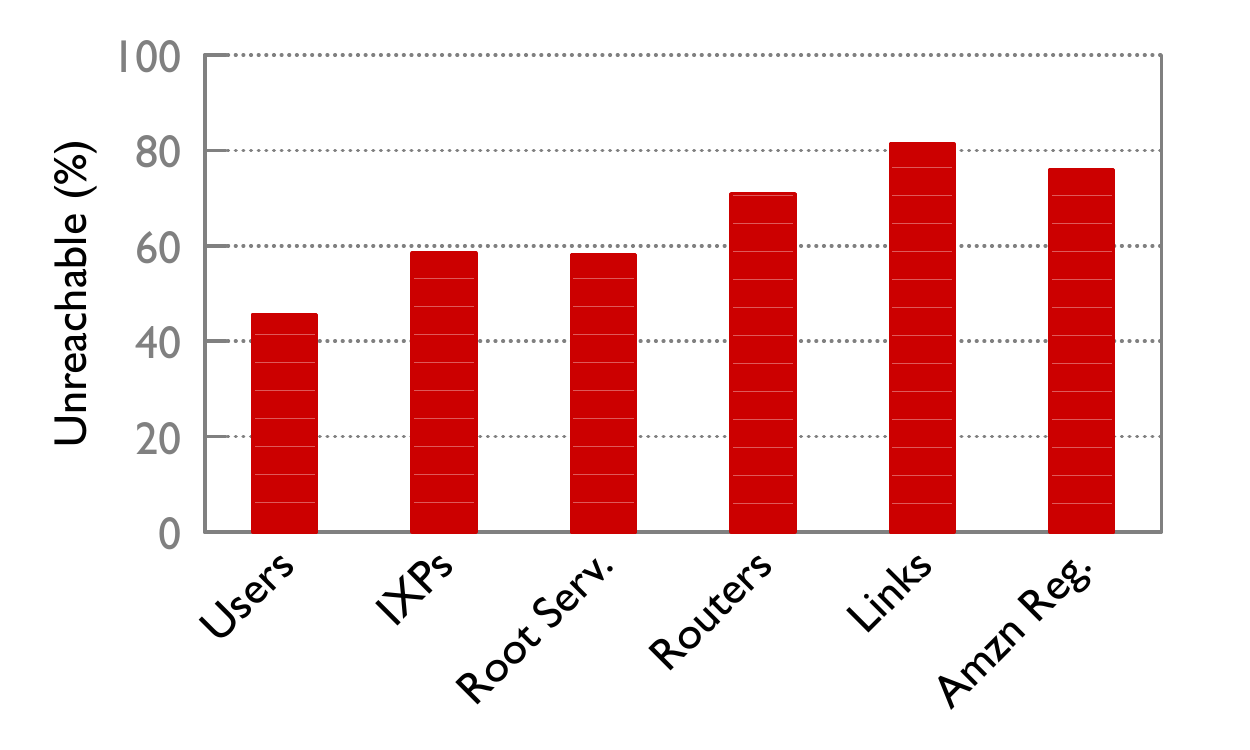}
    \label{fig:failure-eu}
    }
    \subfigure[Latitude Threshold of $50^{o}$]{
    \centering
    \includegraphics[width=0.45\textwidth]{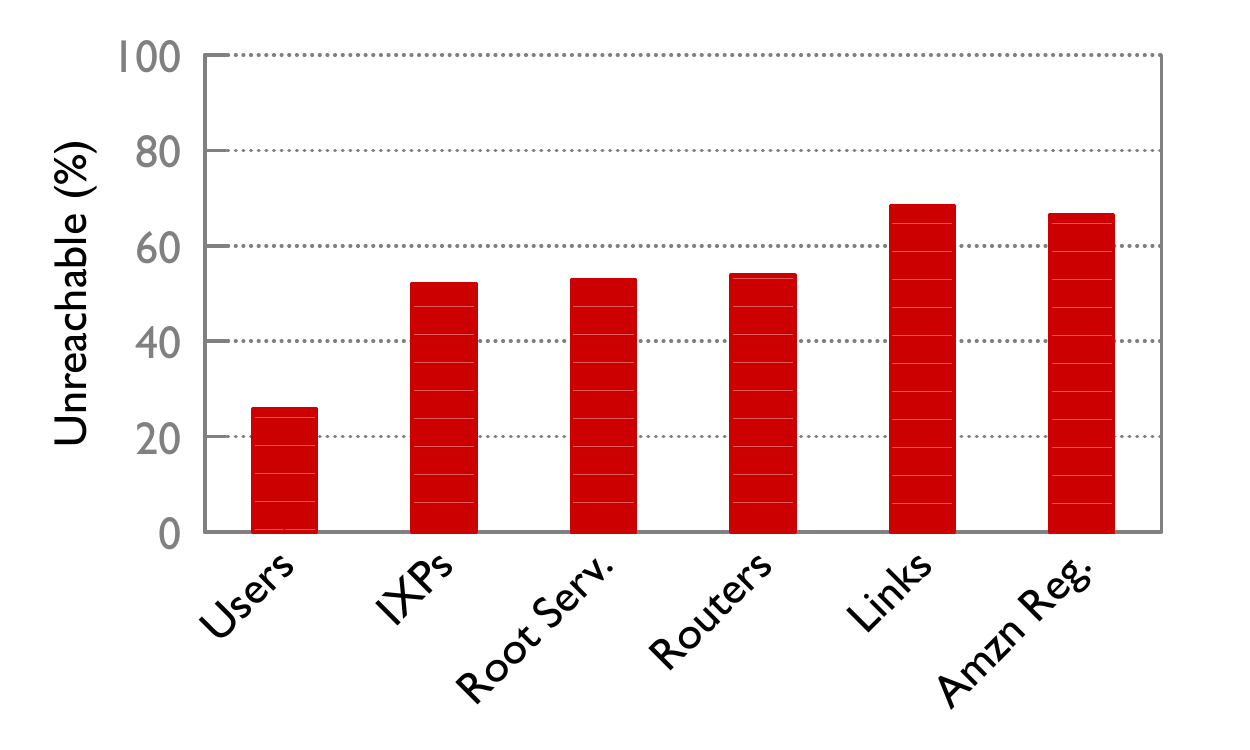}
    \label{fig:failure-na-e}
    }
\caption{Percentage of users and Internet infrastructure components that become unreachable under solar storm impact at different latitude thresholds for storm impact. }
 \label{fig:failure-ss}
\end{figure}

\subsection{Global Failures: Solar Superstorms} 

Solar superstorms~\cite{cisa-1,lloyds} are capable of causing widespread power outages at higher latitudes. We study the indirect impact of solar superstorms on Internet infrastructure by considering WASG failures at higher latitudes. We evaluate failures under two scenarios: using a latitude threshold of $40^{o}$ used in recent work~\cite{sigcomm21} and a relaxed threshold of $50^{o}$. Under each scenario, we consider that WASGs with partial or complete overlap with the vulnerable region (for example, above $40^{o}$N and below $40^{o}$S latitudes for $40^{o}$ threshold) have failed. $20$ WASGs are affected when using $40^{o}$ threshold and $10$ WASGs with a $50^{o}$ threshold.

We repeat the same evaluation as with regional failures. In Figure~\ref{fig:failure-ss}, we observe that a much larger fraction of Internet infrastructure components are expected to become unreachable if WASGs with a presence in higher latitudes fail. Note that the fraction of infrastructure components that were estimated to be unreachable by prior work~\cite{sigcomm21} using a threshold of $40^{o}$ was around $40\%$ for most Internet components. However, the prior work estimated failure rate based on the geolocation of infrastructure components alone. When the geographic spread of WASGs is taken into account, a much larger area is susceptible to failures. 

Many WASGs with a presence above the latitude threshold extend much farther toward the equator. For example, most parts of California that house several Internet components are below the $40^{o}$ threshold and were not considered to be vulnerable by prior work~\cite{sigcomm21}. However, since this region is powered by the Western Interconnection, which has a significant presence in higher latitudes, California is susceptible to cascading grid failures. The geographic span of WASGs extends the vulnerability region for a given threshold to regions far beyond the threshold.

\subsection{Loss of Connectivity Analysis}
\begin{figure}[h]
	\centering
	\includegraphics[width=0.6\linewidth]{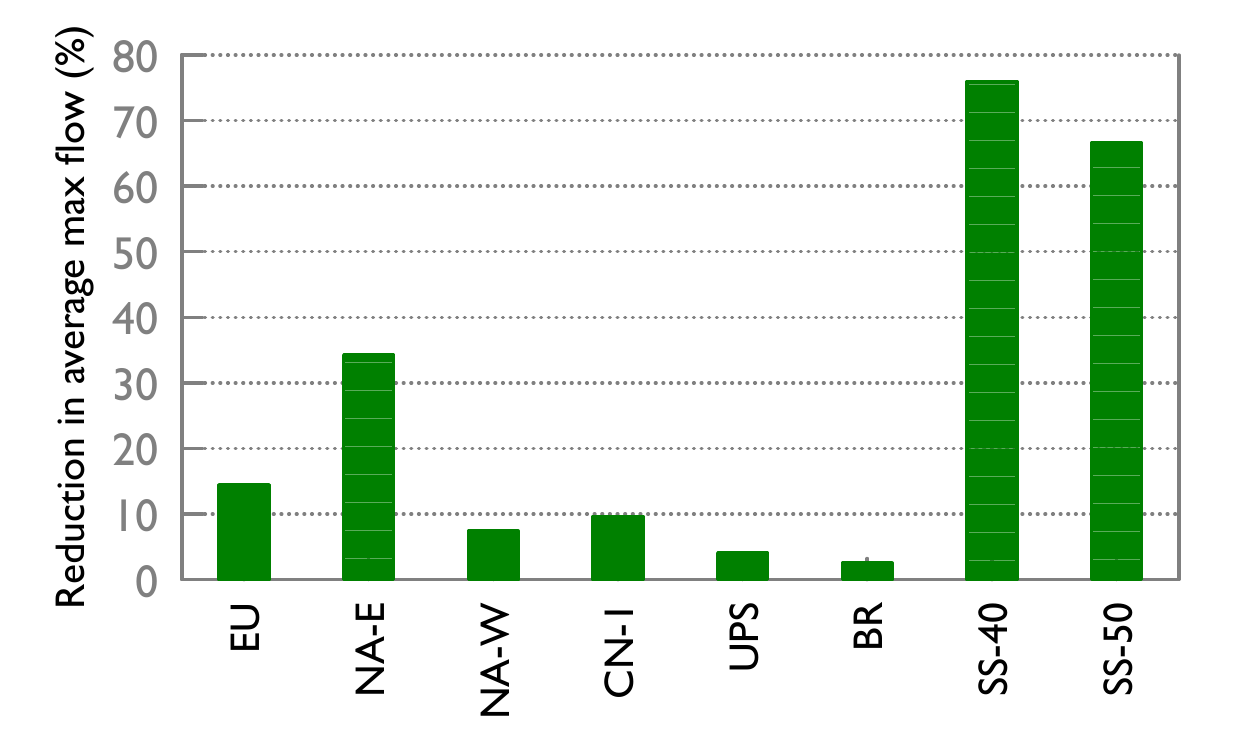}
	\caption{\textbf{Connectivity Loss:} Reduction in average max. flow on the WASG graph with link capacity as the number of links during failures of various regions and solar storm-induced failures at $40^{o}$ and $50^{o}$ thresholds.}
	\label{fig:failure-flows}
\end{figure}

To understand the extent of loss of connectivity in the Internet topology due to failures of WASGs, we conduct maximum flow measurements in a graph where the nodes denote WASGs and capacities on links denote the number of IP links in the Internet topology between the endpoint WASGs of that link. We measure the maximum flow between all pairs of nodes in this WASG graph before and after failures. For each failure scenario, we calculate the reduction in maximum flow for all pairs of nodes and report the average reduction in Figure~\ref{fig:failure-flows}. We present results for the six regional scenarios and two solar storm scenarios with thresholds $40^{o}$ (SS-40) and  $50^{o}$ (SS-50). 

Among regional failure scenarios, we observe that NA-E failure will lead to the largest reduction in maximum flow in the WASG graph. Although the fraction of links that are unreachable during regional failures was only slightly higher for NA-E compared to EU (Figures~\ref{fig:failure-na-e} and \ref{fig:failure-eu}), the overall reduction in maximum flow is more than $2\times$ for NA-E. This shows that NA-E is more critical for Internet connectivity between a larger fraction of other WASGs. While Great Britain, Ireland, and Nordic countries offer paths between North America and Africa/Asia during EU failure, NA-E failure will result in a near complete drop in cross-Atlantic paths. 

Solar storm-induced failures across multiple WASGs show a significant drop in maximum flow. SS-50 is only slightly better than SS-40. This is because the few highly connected WASGs, EU, NA-E, and NA-W, are vulnerable even under a relaxed threshold. A large fraction of East-West paths is affected when these three WASGs fail.

\vspace{2mm}
\parab{Summary} \\
Our analysis shows that:
\begin{itemize}[leftmargin=*,itemsep=1pt]
    \item Nearly $65\%$ of Internet infrastructure components are concentrated in fewer than 10 WASGs (Figure~\ref{fig:wasg-stats}).
    \item NA-E and EU WASGs host the largest fraction of Internet infrastructure components (Figure~\ref{fig:top15}).
    \item Regional failures in NA-E or EU can leave nearly $10-20\%$ of the infrastructure unreachable (Figure~\ref{fig:top15}).
    \item Multiple availability zones and regions of cloud providers that share the same WASG and are susceptible to joint failures during grid outages (Figure~\ref{fig:top15amazon}). When power grid dependence is taken into account, the number of isolated Amazon Availability Zone decreases from 87 to 19.
    \item Risks posed by solar storms could be worse than the estimates in prior work~\cite{sigcomm21} since WASGs in higher latitudes extend further south (Figure~\ref{fig:failure-ss}).
\end{itemize}

\section{Resilience Analysis Tool}
In this section, we provide an overview of NetWattZap, a resilience analysis tool, and demonstrate its use cases.

\subsection{NetWattZap Design}
NetWattZap is a resilience analysis tool with the primary goal of predicting Internet infrastructure or application component deployment strategies that satisfy users' resilience goals with respect to power grid failure zones. Alongside power grid dependency, it also  offers support for other constraints, including latency, location preferences, and cost models. In addition, NetWattZap can be extended to any user-defined requirements that can be represented using linear constraints. 

NetWattZap takes as input user goals and generates an Integer Linear Program to solve the placement problem. The tool internally maintains the power grid infrastructure map with the WASG boundaries and various Internet infrastructure component maps, such as data center locations and IXP locations. NetWattZap generates the constraints of the optimization problem while taking into account various infrastructure dependencies. It then solves the ILP by using the Gurobi solver~\cite{gurobi}.

We present a list of constraints NetWattZap supports. Note that the target infrastructure on which the user specifies the constraints may be data center locations, IXP locations, user-specified private locations, etc. The component being placed may be an application instance, AS presence at an IXP, etc. The constraints below can be specified for any components and targets.

\noindent
\textbf{Grid Resiliency: } The user can specify whether a component requires fully or partially disjoint target locations. Some examples: (i) an application should have instances in three data centers located on different WASGs, (ii) at most two of four data center deployments can be in one WASG.

\noindent
\textbf{Location: } The user can explicitly specify geographic constraints. For example, an application should have at least one instance in the US and in the Southern Hemisphere.

\noindent
\textbf{Latency: } The user can specify latency constraints from a single source or a distribution of sources. The user may also specify overall latency minimization as an objective. The user may provide latencies based on measurements. In our experiments, we use Haversine distance and speed of light in fiber to estimate the latency between any two locations.

\noindent
\textbf{Cost: } The user may specify cost constraints for their deployment. Since NetWattZap only holds location information currently, the user also needs to provide an associated cost model. For example, if the user provides the cost of deployment at various data center locations, NetWattZap can generate an ILP that minimizes the overall cost.

A user's requirements might include a combination of constraints. Note that this list is non-exhaustive, and any user-defined linear constraint may be added to NetWattZap.

A simple application of NetWattZap for selecting data centers for application deployment is shown below. The target infrastructure is the set of data centers $D = \{1,2,...,d\}$. $G[d]$ denotes the ID of WASG that hosts data center $d$. The application has a user population across multiple locations, $U = \{1,2,...,u\}$, with the number of users at a given location, $j$, specified as users[$j$]. If the application developer wants to deploy an application instance across $N$ data centers with at most one data center located in any WASG, the ILP  generated is given below. $x[i]$ is a binary variable that specifies whether a data center is selected. Constraint~\ref{eq:constraint1} enforces that at most $N$ data centers are selected.  Constraint~\ref{eq:constraint2} enforces the rule that at most one data center can be selected from any WASG region. The objective minimizes the overall latency experienced by users.

\begin{align}
\text{Objective:} \quad & \min \sum_{i=1}^{d}\sum_{j=1}^{u} \text{users}[j] \cdot \text{distance}(i, j) \cdot x[i] \label{eq:objective} \\
\text{subject to:} \quad & \sum_{i=1}^{d} x[i] = N \quad \forall i \in D \label{eq:constraint1} \\
& x[i] + x[j] \leq 1 \quad \forall (i, j) \in D, \text{G}[i] = \text{G}[j] \label{eq:constraint2}
\end{align}

\subsection{NetWattZap Use Cases}
In this section, we present the results of two use cases with an additional use case presented in Appendix~\ref{sec:netwattzap3}.

\begin{figure}[ht!]
\centering
    \subfigure[ Multi-User Application Deployment]{
    \centering
    \includegraphics[width=0.75\textwidth]{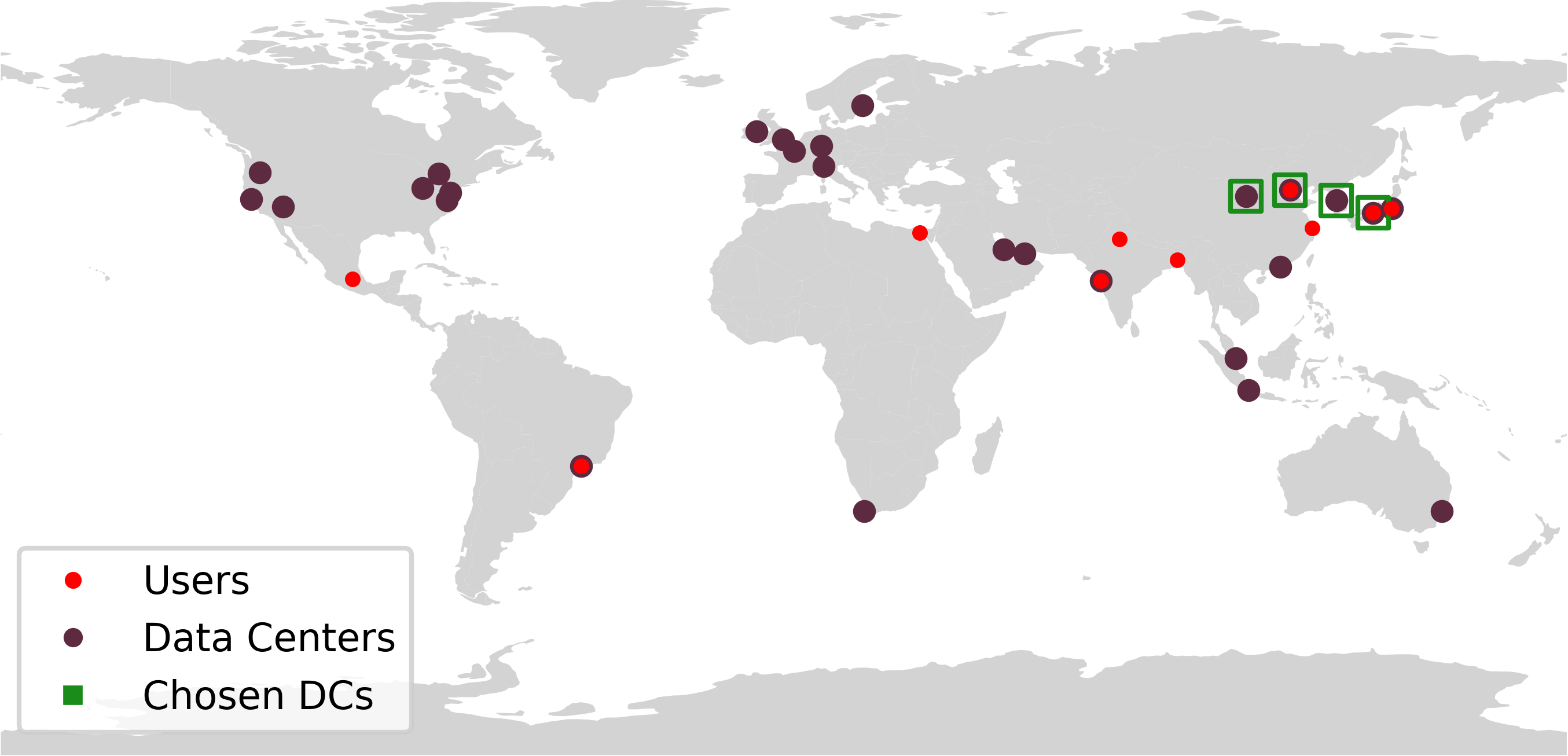}
    \label{fig:multiUserDC}
    }
    \subfigure[ IXP selection]{
    \centering
    \includegraphics[width=0.75\textwidth]{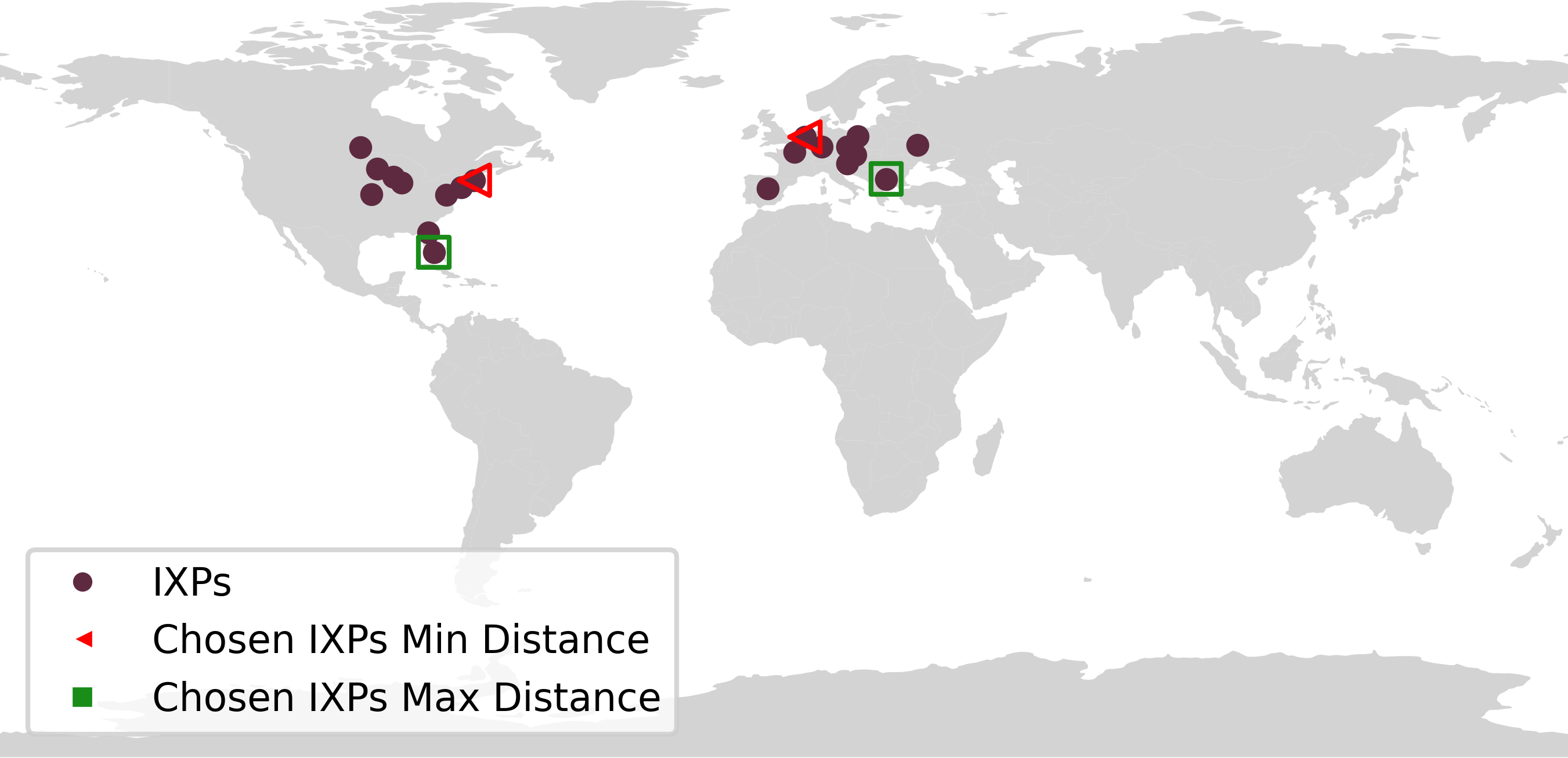}
    \label{fig:ixp-map}
    }
\caption{NetWattZap use cases: (a) An application with users across multiple cities with  grid resiliency and latency constraints. (b) Two ASes with a presence in multiple IXPs pick two IXPs  for peering with grid resiliency at min/max distance. (For visual clarity, WASGs are not shown.)}
 \label{fig:WattZap-uses}
\end{figure}

\parab{Multi-user application deployment: } We consider an application with users in the top ten most populated cities, with a user base proportional to its population. The goal is to find four data centers in disjoint WASGs while minimizing overall user latency. Since the most populated cities are concentrated in Asia, NetWattZap identifies four data centers in Asia, all located in different WASGs (Figure~\ref{fig:multiUserDC}). Note that this is an extended version of the sample LP in the previous subsection with additional constraints to ensure that latency from each city to only the nearest selected data center is considered using additional continuous decision variables.

\parab{IXP selection: } Two Autonomous Systems (ASes) with a presence in multiple IXPs use NetWattZapp to select two IXPs to peer at based on grid resiliency and latency constraints. Chosen IXPs under two different requirements---minimizing/maximizing the distance between chosen IXPs---are shown in Figure~\ref{fig:ixp-map}.  The maximum distance requirement, for example, ensures that the chosen IXPs are farthest away from each other while conforming to disjoint WASG requirements. 

Appendix~\ref{sec:netwattzap3} shows a third use case where a user wants to pick $N$ closest data centers with grid resiliency while minimizing latency. Latency is estimated using Haversine distance in all examples, but may be replaced with user-provided latency measurements in practice.
\section{Discussion}
 Today's society is a system of systems in which failure of one part could have disastrous consequences for the entire system~\cite{global-risk-1, multi-1, multi-2}. Since the Internet constitutes a crucial part of this interconnection, it is necessary to not only improve the resilience of the Internet independently but also in conjunction with other infrastructures. In this section, we discuss some open questions and research directions toward improving the long-term resilience of critical infrastructures.

\parab{Improving Internet and Cloud Resilience:} Our preliminary study has revealed several blind spots in our understanding of Internet resilience. We surmise that the organic growth of Internet infrastructure and even the well-planned expansion of large cloud companies have overlooked the power grid failure zones based on our findings. Deployment of Internet infrastructure components (for example, new anycast servers, IXPs, etc.) should take into account the redundancy characteristics under WASG failures. Deployment of distributed applications in the cloud can also benefit from power dependency awareness with NetWattZap.

\parab{A Deeper Dive into Power and Internet Interdependency Analysis:} The characterization of Internet's dependence on power grids can be improved on many fronts: \textit{(i) More accurate Internet maps:} We need better mapping with improved coverage of Internet infrastructure component locations and their interconnections. Current maps only have partial information about the global infrastructure. \textit{(ii) Backup power information: } A dataset with backup power duration available for various Internet components can further strengthen the dependence analysis. \textit{(iii) Better WASG failure modeling: } Since WASG failure modeling is a topic of active research~\cite{topo-impact,topo-impact2,topo-impact3,cascade-model1,cascade-model2,cascade-model3,cascade-model4,cascade-model5,cascade-model6,cascade-model7,cascade-model8,cascade-model9}, cross-infrastructure dependence analysis can leverage better models developed by the power grid community in the future. \textit{(iii) Non-WASG grid failure modeling: } On the power infrastructure front, characterizing failures in regions that are not covered by WASGs can also be helpful for understanding the impact on the Internet. 

Our analysis framework can be extended in the future with fine-grained failure dynamics of both the Internet and power infrastructures as our understanding of both these domains evolves. 

\parab{Extending to Other Infrastructures:} Beyond the Internet and power grids, it is also necessary to extend the interdependency analysis to other infrastructures. Several past works that analyze a network of networks~\cite{multi-1,multi-2, multi-3} rely on network properties such as node importance and node centrality to characterize the complexity. However, our preliminary analysis shows that knowledge of the network topology alone is not sufficient. Our critical infrastructures are complex systems with different operational characteristics and failure patterns. For example, the notion of a well-bounded region of failure with WASGs in power grids is radically different from routing in the Internet, which adapts to available network topology. Hence, the study of interconnections of operational networks cannot be mere graph analysis at a larger scale. Understanding the dynamics of each network and its interplay is crucial. This is an opportunity for Internet researchers to be actively involved in such cross-disciplinary research. Our analysis framework can also be used by other domains that rely on power grids and the Internet. For example, airlines, logistics, and other globally distributed service domains can use our resilience analysis while deploying their hubs to improve overall resilience and service availability. 

\parab{Renewable Energy Sources:} Grids around the globe are moving towards renewable energy sources and increased decarbonization. The resilience characteristics of grids are evolving as the nature of energy sources changes. In this evolving context, it is worth noting that while the probabilities of component failures can change, the fundamental issue of shared failure zone for the distribution infrastructure remains. WASG failures can happen irrespective of the nature of the energy source used in power plants. Only local power generation, such as solar on rooftops, is immune to WASG failures. Power generated in a remote solar farm and then transported to homes are also vulnerable to WASG failures. 

It is also worth noting that while repeaters in submarine cables are powered by power-feeding equipment at either end of the cable and a conductor running through the length of the cable, terrestrial cables have a more simple electrical configuration~\cite{power-cable}. Repeaters in terrestrial long-haul cables are powered by local sources. Conversations with cable owners also confirmed that repeaters are powered in ``huts'' located along the length of the cable which draws power from the local grid supply. Hence, even if data centers have local renewable power sources, links connecting to them can be affected by WASG failures.

\section{Conclusion}
The Internet and power grids constitute the backbone infrastructures of our society. It is imperative to study the Internet's dependence on the power infrastructure to discern the nature of threats that the communication infrastructure might encounter. This paper takes the first step in this direction. Our analysis shows that a large fraction of Internet infrastructure components is concentrated in a few power grid failure zones. Moreover, the current notion of cloud availability regions may not hold ground when power grid failure zones are taken into account. Our study underscores the need for an in-depth analysis of the interdependencies of the Internet with other infrastructures. 

\noindent
This work does not raise any ethical concerns.

\clearpage

\bibliographystyle{ACM-Reference-Format}
\bibliography{grid-internet}

\appendix

\section{Wide Area Synchronous Grids}
\label{sec:wasg}
In this section, we provide a brief overview of Wide Area Synchronous Grids across the globe since they form the basis for understanding the resilience of the Internet in relation to power grids~\footnote{Note that the boundaries are drawn at the granularity of states in Figure~\ref{fig:wasg-global}. The borders between WASGs do not strictly follow state boundaries in a small number of states in the US and Japan, although the difference of a few counties may be too small to be discernible at the plotted scale. Accurate data is included in the dataset}.

\parab{North and Central America: }
There are two major and several smaller interconnections in North America. The Eastern Interconnection (\textbf{NA-E}~\footnote{The name in bold denotes the short form used to refer to the grid in tables and plots. These are not standard or official acronyms.}) covers the US states to the east of the Rockies excluding Texas and eastern states of Canada except Quebec~\cite{usgrid-info}. The Western Interconnection (\textbf{NA-W}) spans the western states of the US and Canada~\cite{usgrid-info}. The US states of Texas (\textbf{TX}) and Alaska~\cite{usgrid-info} and the Canadian state of Quebec~\cite{quebec-info} have their own regional grids that are not connected to the major grids.

Apart from a small part of northern Baja California connected to the Western Interconnection, Mexico (\textbf{MX}) runs a synchronized grid across the country~\cite{mexico-info}. SIEPAC (Central American Electrical Interconnection System)~\cite{siepac-info} interconnects the power grids of six Central American countries. 

\parab{South America: } The Argentine Interconnection System~\cite{argentina-info} (\textbf{AR}) covers Argentina and its neighboring countries Uruguay and Paraguay. While there are efforts to synchronize Argentina with Chile, the grids are not synchronized yet~\cite{argentina-chile}. Andean countries of Colombia, Peru, and Ecuador are currently synchronized~\cite{andean} with additional plans to extend the WASG to Bolivia and Venezuela. Brazil (\textbf{BR}) operates a synchronized national grid~\cite{brazil-info}.

\parab{Europe:} The synchronous grid of continental Europe (\textbf{EU}) covers over $25$ countries in Europe and is also interconnected with Turkey and the northern African countries of Morocco, Algeria, and Tunisia~\cite{eu-info}. The Nordic grid interconnects Nordic countries and parts of Denmark. Great Britain~\cite{gb-info} (\textbf{GB}) and Ireland~\cite{ireland-info} operate their own independent national grids. Baltic countries are currently connected to the Russian grid but plan to join the European grid by 2025~\cite{baltic-info}.

\parab{Africa:} Southern African Power Pool \textbf{(SAPP)} comprises of $12$ countries in southern Africa and West African Power Pool \textbf{(WAPP)} spans $14$ countries~\cite{wapp-info}. Egyptian and Libyan grids are interconnected with the Arab grid that connects with Iraq, Jordan, etc.~\cite{arab-info}  (\textbf{Arab-1})

\parab{Asia:} India (\textbf{IN}) integrated its regional grids to form a synchronized national grid spanning the entire country (except its islands)~\cite{india-info}. China has six regional grids of which three (North, Central, and East China grids) are interconnected and synchronized~\cite{china-info} (\textbf{CN-1}). China also has the only known case of splitting up of a synchronous grid---Yunnan province recently disconnected from the Southern grid (\textbf{CN-2}) for better resilience~\cite{china-1}. Pakistan (\textbf{PK}) has a synchronized national grid~\cite{pak-info}. The grids of six Gulf countries are also synchronized~\cite{gcc-info}.

Integrated Power System/Unified Power System (IPS/UPS) that envelopes many former Soviet countries witnessed high churn over the past decade~\cite{russia-info} (\textbf{UPS}). Turkmenistan, Uzbekistan, and Tajikistan left the grid due to political reasons and are now a part of the Central Asian Power System~\cite{caps-info}. Ukraine and Moldova also left UPS this year during the Russian invasion~\cite{russia-info}. Iran recently made plans to join this grid~\cite{iran-info} while the Baltic countries are planning to leave IPS/UPS for the European grid~\cite{baltic-info}.

The synchronized JAMALI system in Indonesia (\textbf{ID}) extends across three islands and handles $75\%$ of the nation's energy sales~\cite{indonesia-info}. North Korea and South Korea (\textbf{KR}) operate their own grids which are not interconnected~\cite{sKorea-info}. Japan is the only country with two grids---Eastern (\textbf{JP-1}) and Western (\textbf{JP-2})---that is operating at two different frequencies due to historical reasons, which makes inter-operation and synchronization very difficult~\cite{jp-info}.

\parab{Oceania: } Australia has two large synchronized grids and several smaller grids~\cite{aus-info}. The largest WASG in Australia covers the eastern and southeastern states~\cite{aus-info}. New Zealand has a national grid with synchronization within grids of the Northern and Southern Islands~\cite{nz-info}.

\section{Brazil Outage Aug 2023}
\label{sec:br-real}
\begin{figure*}[h]
	\centering
	\includegraphics[width=\linewidth]{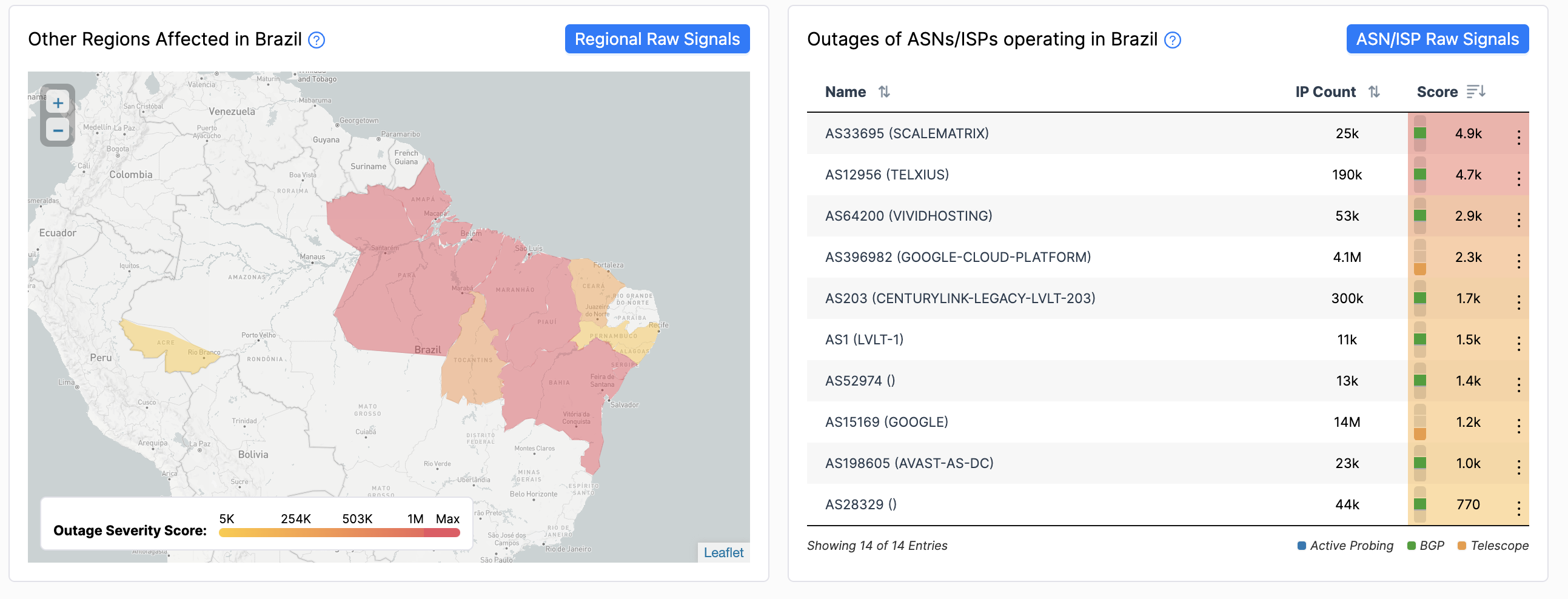}
	\caption{Internet Failure Map during Brazil Power Grid Outage on 15 Aug 2023 from IODA~\cite{IODA} (left). The list of top ASes affected is shown on the right. 
}
	\label{fig:br-failure-real}
\end{figure*}

Figure~\ref{fig:br-failure-real} shows the network outage map during Brazil grid failure~\cite{br-real-2023} on 15 Aug 2023 obtained from the Internet Outage Detection and Analysis (IODA) website~\cite{IODA}. Internet connectivity was affected in a large region. However, it was not a complete connectivity blackout. Note that the scores shown are IODA's outage severity score which is a complex metric that takes into account a variety of factors. 

\clearpage
\section{NetWattZap Use Case 3}
\label{sec:netwattzap3}

\begin{figure*}[h]
	\centering
	\includegraphics[width=0.8\linewidth]{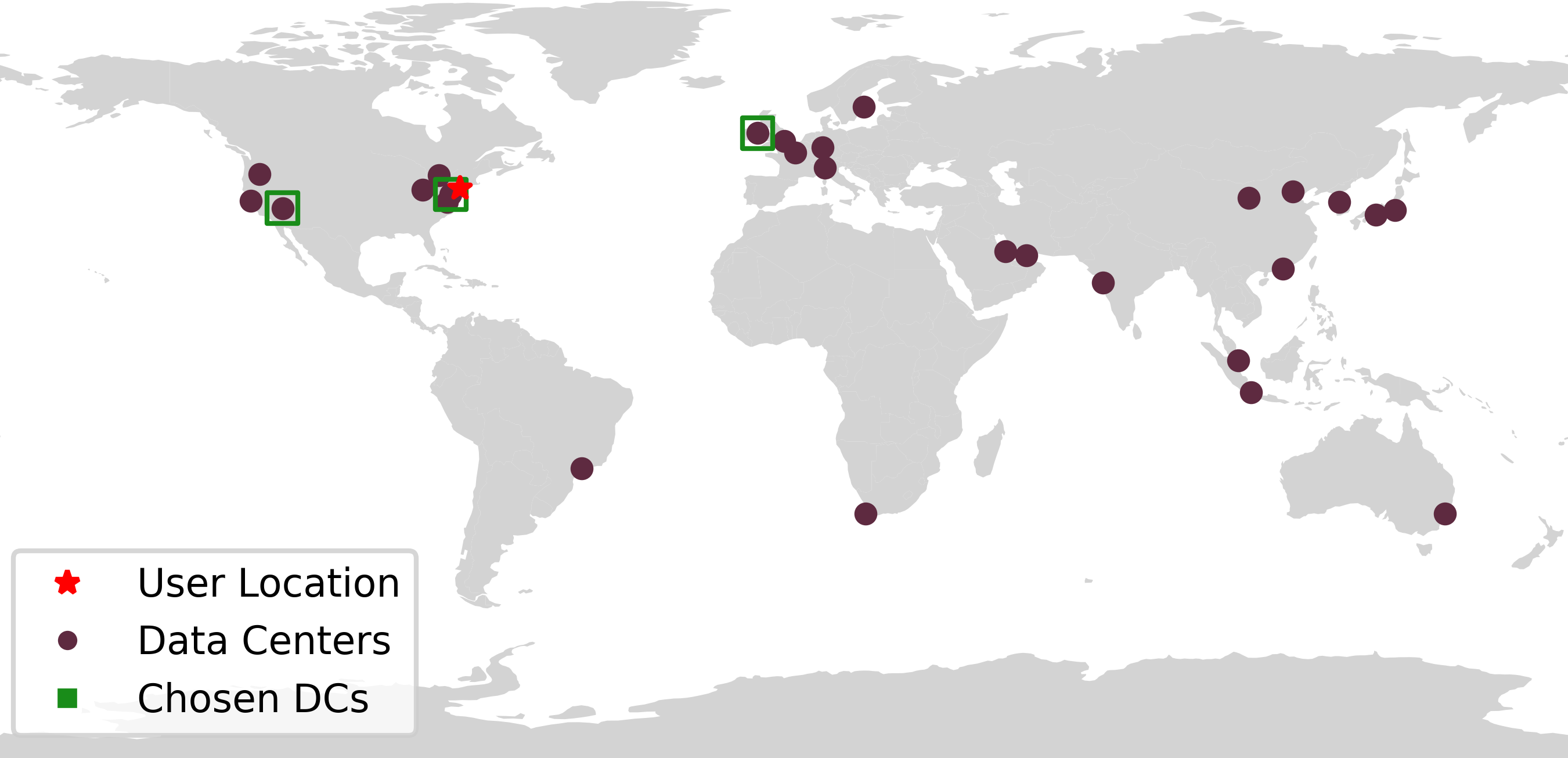}
	\caption{NetWattZap output for a single user in NYC who wants to choose data centers with grid resiliency while minimizing overall latency. 
}
	\label{fig:singleUserDC}
 \vspace{4mm}
\end{figure*}

\parab{Single user data center selection: }
In this example, a single user wants to find three data centers in disjoint WASGs while minimizing the overall latency from the user's location (New York City). The optimization problem identifies a data center each in US Eastern Interconnection, US Western Interconnection, and Ireland, which minimizes the overall latency while being in disjoint WASGs (Figure~\ref{fig:singleUserDC}).

\begin{figure*}[h]
	\centering
	\includegraphics[width=0.9\linewidth]{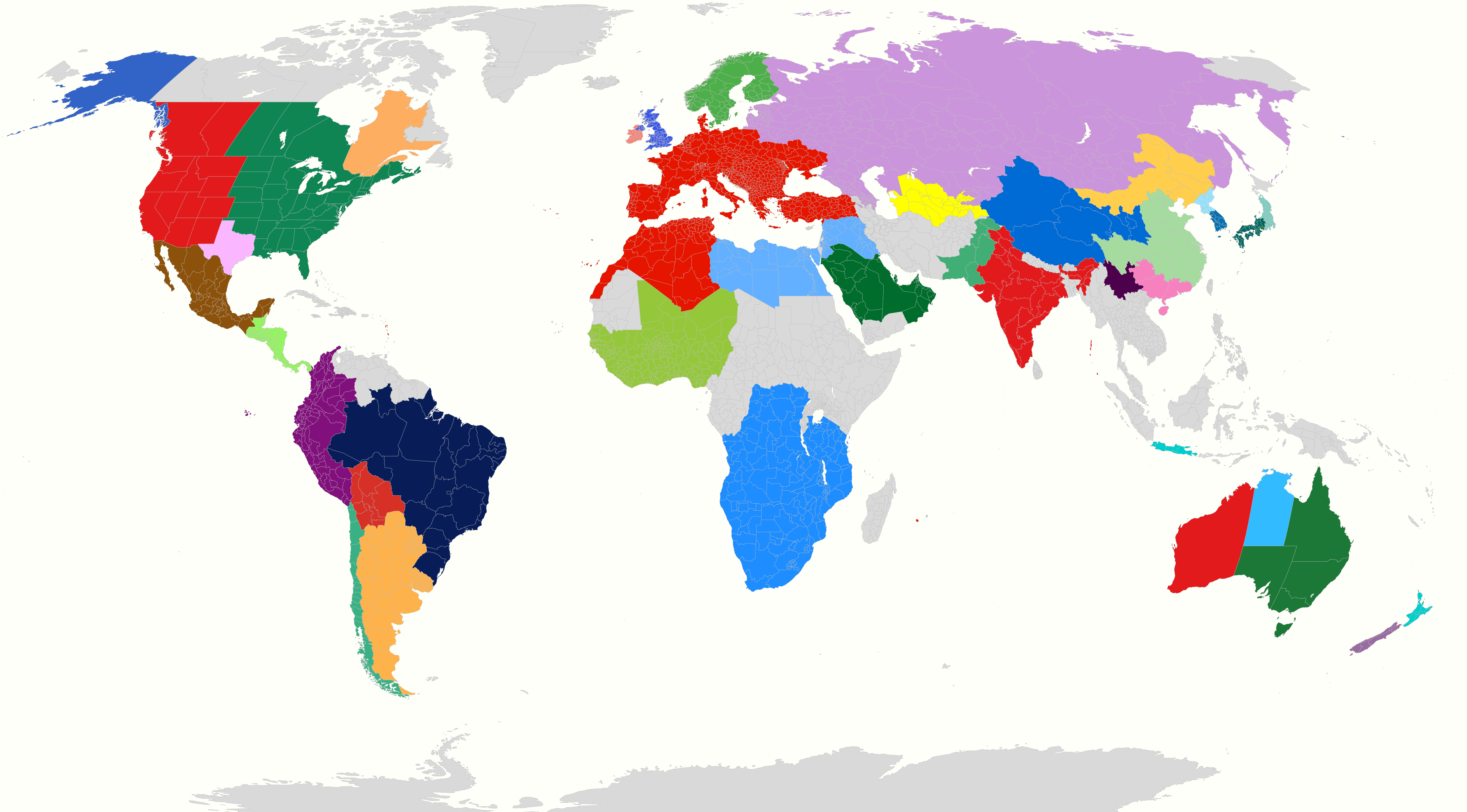}
	\caption{Map of 43 WASGs for visual comparison with Figure~\ref{fig:singleUserDC}. 
}
	\label{fig:wasg2}
\end{figure*}

\end{document}